\documentclass[a4paper,11pt,nofootinbib,floatfix]{revtex4-2}

\usepackage{amssymb,bm,epsf}
\usepackage{setspace}
\usepackage{amsmath}
\usepackage{graphicx}
\usepackage{dcolumn}
\usepackage{color}
\usepackage{accents}
\usepackage{multirow}
\usepackage{tikz}
\usepackage{comment}
\usepackage{subfigure}
\usepackage{soul}
\usepackage[rightcaption]{sidecap}
\sidecaptionvpos{figure}{t}
\usepackage{enumerate}

\usepackage[toc,page]{appendix}
\usepackage[normalem]{ulem}

\usepackage[hyperfootnotes=false]{hyperref}
\hypersetup{
    colorlinks=true,
    linkcolor=red,
    filecolor=magenta,      
    citecolor=blue
}

\usepackage{titlesec}

\titleformat{\section}
    {\normalfont\bfseries\normalsize\centering}
    {\thesection}{1em}{\MakeUppercase}
\titlespacing{\section}
    {0pt}   
    {2.0ex plus .5ex minus .2ex}  
    {0.8ex plus .1ex}             

\titleformat{\subsection}
    {\normalfont\bfseries\centering}{\makebox[1cm][l]{\thesubsection}}{0pt}{}
\titlespacing{\subsection}
    {0pt}
    {1.5ex plus .2ex minus .2ex}
    {0.8ex plus .1ex}

\newcolumntype{P}[1]{>{\centering\arraybackslash}p{#1}}
\newcolumntype{M}[1]{>{\centering\arraybackslash}m{#1}}

\makeatletter
\@addtoreset{equation}{section}

\makeatother

\begin{document}
\begin{titlepage}
\title{Gravitationally Induced UV Completion of an \texorpdfstring{$O(N)$}{O(N)} Scalar Theory}
\author{Alfio M. Bonanno}
\email{alfio.bonanno@inaf.it}
\affiliation{
INAF Osservatorio Astrofisico di Catania, Via S.Sofia 78, 95123 Catania ITALY
\\
and 
INFN, Sezione di Catania, Italy
}
\author{Emiliano Glaviano}
\email{emiliano.glaviano@inaf.it}
\affiliation{
INAF Osservatorio Astrofisico di Catania, Via S.Sofia 78, 95123 Catania ITALY
\\
and 
INFN, Sezione di Catania, Italy
}


\begin{abstract}
We investigate the ultraviolet completion of an $O(N)$ scalar field theory non-minimally coupled to gravity using the Wilsonian functional renormalization group in the proper-time formulation. Focusing on the spontaneously broken phase, we study the RG flow of the scalar potential and the non-minimal curvature coupling expanded around a running minimum. We identify two distinct classes of fixed-point solutions, one of which is ultraviolet attractive and characterized by a vanishing quartic coupling together with finite, interacting gravitational couplings. For a finite region of infrared initial conditions, the RG trajectories remain regular at all scales and approach this fixed point.  This mechanism renders the theory asymptotically safe and leads to a flat scalar potential in the ultraviolet. We show that this mechanism is robust under changes of cutoff scheme and truncation, allowing the ultraviolet completion requirement to constrain the infrared values of the scalar couplings and the mass scale in the broken phase.
\end{abstract}

\maketitle

\end{titlepage}
\newpage
\setcounter{page}{2}

\section{Introduction}\label{introduction}
\noindent Quantum scalar field theories in four dimensions generically face the Landau-pole problem associated with the running of the quartic self-interaction $\lambda$. In practice, a Landau pole signals that the low-energy description cannot be extrapolated indefinitely and that new degrees of freedom or a non-perturbative ultraviolet (UV) completion must enter at sufficiently high scales. In the Standard Model the perturbative running of the Higgs quartic coupling is substantially altered by the top-quark Yukawa interaction, which significantly modifies the ultraviolet behavior of the running and removes the Landau pole. However, this does not generically resolve the ultraviolet problem because $\lambda$ turns negative around an RG scale of $10^{10}\,\mathrm{GeV}$ and subsequently evolves slowly \cite{Buttazzo:2013uya}. In the absence of an UV completion, this leads to an instability of the effective potential, rendering it unbounded from below at high energies.

In the asymptotic-safety scenario for gravity--matter systems \cite{Pastor-Gutierrez:2022nki,Eichhorn:2018whv,Garces:2025rgn} and, in particular in the context of the Standard Model, gravitational interactions, along with Yukawa interactions, provide an UV completion by an UV attractive fixed-point. At high scales, the interplay between quantum gravity and Top fluctuations drive the running couplings in this regime and the Higgs quartic coupling to zero motivating the so-called \emph{flatland} scenario \cite{Shaposhnikov:2009pv,Wetterich:2016uxm,Hamada:2017rvn,Eichhorn:2017ylw,Eichhorn:2018whv,Pawlowski:2018ixd,Wetterich:2019zdo}, building on earlier ideas \cite{Hashimoto:2013hta,Hashimoto:2014ela}; see \cite{Eichhorn:2022jqj,Eichhorn:2022gku} for recent reviews. Related evidence that quantum gravity can cure Landau-pole behavior have been found, for instance, in studies of the Abelian gauge sector \cite{Eichhorn:2019yzm,Eichhorn:2017muy,Eichhorn:2017lry}.

Nevertheless, additional scalar fields beyond the Higgs are widely expected. They may play a role in inflation \cite{Starobinsky:1980te,Guth:1980zm,Mukhanov:1981xt,Linde:1983gd,Shafi:1983hj,Bezrukov:2007ep,Herranen:2014cua}, dark energy \cite{Li:2012dt,Oks:2021hef}, or dark matter \cite{Cirelli:2024ssz,Arbey:2021gdg}, and in such sectors the Standard-Model Yukawa couplings to fermions may be absent or highly suppressed. It is therefore natural to ask whether \emph{gravity alone}, coupled to scalars, can provide a mechanism that renders the theory UV finite, or whether fermionic interactions along with gravity are essential as in the Standard Model.

Scalar--tensor theories offer a minimal setting to address this question \cite{Christiansen:2017cxa,Burger:2019upn,Eichhorn:2018nda,Alkofer:2018fxj,Biemans:2017zca,Oda:2015sma,Meibohm:2015twa,Dona:2013qba}. A particularly interesting case arises when scalars are non-minimally coupled through a term $\xi_{ij}\phi_i\phi_j R$, or more generally via $F(\phi)R$, which ties the running of the non-minimal coupling to the running of the scalar potential in the functional renormalization group (FRG) framework \cite{Wetterich:2022brb,Pawlowski:2018ixd,Wetterich:2019rsn,Wetterich:2019zdo}. In the symmetry-broken phase, the presence of $\xi_{ij}$ can have important implications at scales above the Planck mass $m_p$ \cite{Shapiro:2015ova,Merzlikin:2017zan,Shaposhnikov:2009pv,Wetterich:2019rsn,Pawlowski:2018ixd,Wetterich:2022brb,Wetterich:2019zdo}. The RG flow of $\xi$ has been studied for $N=1$ in \cite{Shapiro:2015ova} and for general $F(\phi)R+U(\phi)$ systems in \cite{Merzlikin:2017zan} using Wetterich-type flows \cite{Morris:1993qb,WETTERICH199390,Ellwanger:1993mw}.
However, those investigations focused on the infrared (IR) regime, since the gravitational part of the Hessian was neglected and only scalar fluctuations were retained. In contrast, in the UV regime $m_p^2\ll p^2$ gravitational fluctuations can dominate the dynamics, especially if gravity itself exhibits an interacting fixed point that can substantially alter the flow in the high energy limit. 

In recent years, functional renormalization group (FRG) techniques have provided substantial evidence for asymptotic safety in gravity. In particular, using the linear split $g_{\mu\nu}=\bar g_{\mu\nu}+h_{\mu\nu}$ and extensions of the Einstein--Hilbert truncation including $f(R)$, $f(R_{\mu\nu}R^{\mu\nu})$ and $f(R_{\alpha\beta\mu\nu}R^{\alpha\beta\mu\nu})$-type truncations \cite{Machado:2007ea,Codello:2008vh,Falls:2013bv,Falls:2014tra,Falls:2018ylp,Codello:2007bd,Falls:2017lst}, a non-Gaussian UV fixed point has been consistently observed. These studies indicate a small number of relevant directions, with evidence that higher-curvature operators do not introduce additional relevant couplings within the explored truncations. In particular, a partial convergence of universal quantities, such as critical exponents, is also observed as the truncation is systematically enlarged. In addition, the inclusion of higher-order contributions, such as the two-loop counterterm of Goroff and Sagnotti \cite{GOROFF1986709}, has been shown not to spoil the existence of the UV fixed point, and the corresponding operator appears to be irrelevant \cite{Gies:2016con}. In contrast, analyses based on $f(R)$-type truncations combined with the exponential parametrization $g_{\mu\nu}=g_{\mu\alpha}(e^h)^\alpha_{\ \nu}$ \cite{Alkofer:2018fxj,Ohta:2015fcu,Ohta:2015efa,Falls:2016msz,Ohta:2016npm,Goncalves:2017jxq,Nink:2014yya,Falls:2015qga,Gies:2015tca} typically find a different number of relevant directions, suggesting a sensitivity of quantitative results to the choice of parametrization. This issue has been further analyzed in \cite{DeBrito:2018hur}, where aspects of this discrepancy have been clarified.

The existence of a UV-attractive fixed point persists also in gravity--matter systems, both in the Einstein--Hilbert truncation with linear parametrization \cite{Percacci:2003jz,Narain:2009fy,Narain:2009gb,Dona:2013qba,Meibohm:2015twa,Christiansen:2017cxa,Litim:2018pxe} and in $f(R)$-type extensions \cite{Benedetti:2009gn,Hamada:2017rvn,Biemans:2017zca,Alkofer:2018fxj}. Similar results are obtained within the exponential parametrization as well \cite{Kawai:1992np,Percacci:2015wwa,Labus:2015ska,Bonanno:2025qsc}, although quantitative features may depend on the gauge choice \cite{Ohta:2021bkc}. The dependence of fixed-point properties on gauge and parametrization is a well-known issue in asymptotically safe gravity \cite{Falkenberg:1996bq,Souma:2000vs,Barra:2019rhz,Bonanno:2025tfj,Ohta:2016npm,Ohta:2016jvw}. This becomes more involved in gravity--matter systems for two main reasons: first, different gauge choices modify the projection of metric fluctuations onto matter sectors; second, truncation and regularization schemes introduce additional scheme dependencies. Parametrization ambiguities can further contribute to such effects, making it difficult to disentangle genuine physical features from truncation artifacts. At present, there is no unique criterion that selects a preferred parametrization or gauge in the functional renormalization group. In gravity--matter systems with scalar fields, the RG flow equations in the linear parametrization can develop singular behavior at finite values of the couplings within the truncations considered, whereas the exponential parametrization leads to a regular flow in the same regime. In addition, with this parametrization physical quantities in truncated RG flows generally exhibit a residual, typically mild dependence on the choice of gauge and parametrization \cite{Nink:2014yya,Falls:2015qga}. These features motivate the use of the exponential parametrization together with the physical gauge \cite{Percacci:2015wwa,Ohta:2015fcu}. These choices have proven particularly useful in scalar--tensor systems as well \cite{Percacci:2015wwa,Labus:2015ska,Bonanno:2025qsc}.

An interacting UV completion can also make low-energy physics predictive. In the Standard-Model context, the approach to a gravity-induced UV fixed point can correlate IR values of couplings and masses, with early examples including bounds on the Higgs mass \cite{Shaposhnikov:2009pv} and quark-mass relations \cite{Eichhorn:2017ylw,Eichhorn:2018whv}, as well as predictions for Abelian gauge couplings \cite{Eichhorn:2017muy,Eichhorn:2018yfc,Eichhorn:2017lry}. Beyond the Higgs sector, asymptotic safety may also shed light on dark-matter model building \cite{Eichhorn:2017als,Reichert:2019car,Hamada:2020vnf,Eichhorn:2020kca}, where a scalar dark sector could be driven to an interacting UV regime by gravitational fluctuations.

In this paper, we study an $O(N)$ scalar theory non-minimally coupled to gravity of the form $F(\phi)R+U(\phi)$ and investigate whether gravity alone can render the theory UV finite. 
We work in a Wilsonian setting using the proper-time (PT) formulation of the functional RG \cite{Fock:1937dy,PhysRev.82.664,Bonanno:2000yp,Mazza:2001bp,Liao:1994fp,Liao:1995nm,Schaefer:1999em,Schaefer:2001cn,Bohr:2000gp,Zappala:2001nv,Litim:2010tt,Bonanno:2022edf,Glaviano:2024hie,Giacometti:2025qyy,Bonanno:2025tfj}. The PT flow for the Wilsonian action $S_\Lambda$ reads
\begin{equation}
\label{PTFE}
\Lambda\partial_\Lambda S_\Lambda[\phi]= \frac{1}{2}\int_{0}^{\infty}\frac{ds}{s}\,r\!\left(s,\Lambda^2 Z_\Lambda\right)\, \mathrm{Tr}\!\left[e^{-sS_\Lambda^{(2)}}\right],
\end{equation}
where $\Lambda$ is the Wilsonian UV cutoff, $S_\Lambda^{(2)}$ the Hessian, $Z_\Lambda$ the wave-function renormalizations, and $r$ a cutoff kernel. We employ the spectrally adjusted cutoff family discussed in \cite{Bonanno:2019ukb},
\begin{equation}
\label{cutoff}
r(s,\Lambda^2 Z_\Lambda)= \left(2 + \epsilon\frac{\Lambda \partial_\Lambda Z_\Lambda}{Z_\Lambda}\right)\frac{(s\, \gamma\, \Lambda^2 Z_\Lambda)^m}{\Gamma(m)}e^{-s\, \gamma\, \Lambda^2 Z_\Lambda},
\end{equation}
where $m>0$ controls the shape in the interpolation region, $\gamma\in\{m,1\}$ labels two cutoff families, and $\epsilon\in\{0,1\}$ distinguishes type-C ($\epsilon=0$) from type-B ($\epsilon=1$) cutoffs; throughout this work we adopt the type-C choice, which is known to yield highly accurate critical exponents at the Wilson-Fisher fixed point \cite{Mazza:2001bp,Zappala:2001nv} but also in the pure gravitational sector \cite{Bonanno:2004sy}.

Our goal is to determine under which conditions the coupled scalar--gravity system becomes asymptotically safe, thereby providing a concrete realization of a gravity-induced mechanism that renders the scalar sector UV complete. We analyze the fixed-point structure, the critical properties, and the global behavior of RG trajectories in the symmetry-broken phase, identifying a finite region of infrared initial data that flows to a UV-attractive fixed point with vanishing quartic coupling. 

In order to properly interpret the results, it is important to clarify the status of the renormalization group framework employed in this work. The proper-time flow equation does not belong to the class of exact renormalization group equations for the effective average action (EAA). Nevertheless, it has been successfully applied in a wide range of contexts, including the Standard Model \cite{Giacometti:2025qyy}, pure gravity \cite{Bonanno:2025dry,Bonanno:2025tfj,Giacometti:2024qva,Bonanno:2023fij,Bonanno:2023ghc,Bonanno:2012dg}, and gravity--matter systems \cite{Bonanno:2025qsc,Spina:2025wxb}, where it reproduces results consistent with other functional renormalization group approaches. More recently, the proper-time framework has been analyzed within a Wilsonian setting \cite{deAlwis:2017ysy,Bonanno:2019ukb}, where, starting from a UV-regulated Wilsonian action, an exact functional renormalization group scheme based on proper-time regularization can be defined. In the present work, our goal is not to describe the flow of an EAA, but rather to adopt this Wilsonian perspective and use the proper-time flow as a tool to investigate the renormalization group evolution of the action $S_\Lambda = F_\Lambda(\phi) R + U_\Lambda(\phi)$, following the framework developed in \cite{Bonanno:2019ukb}. In this framework, a key ingredient is the existence of a coarse-graining function $\Psi_x^\Lambda$ required in order to define the Wilsonian action. While its explicit construction for interacting theories is highly non-trivial and goes beyond the scope of the present work, we assume that such a function exists $\Psi_x^\Lambda$ for the class of models considered here and proceed to analyze the resulting flow equations within the chosen truncation.

The paper is organized as follows. In Sec.~\ref{sectionformalism} and \ref{sec:approximations} we introduce the model and the discuss in detail the approximation scheme. In Sec.~\ref{sectionbeta} we present the beta functions and discuss their structure. In Sec.~\ref{sectionfixedpoint} we analyze the fixed points and critical exponents. In Sec.~\ref{sectioncriticalline} we explain the basic physical mechanism  and delineate the UV-complete region of initial conditions. In Sec.~\ref{sectionnumsol} we present numerical solutions and discuss the resulting predictions. We summarize our conclusions in Sec.~\ref{sectionconclusion}. Appendix~\ref{appZLZT} provides the derivation of the flow equations for the wave-function renormalizations. Appendix~\ref{appbeta} collects the explicit beta functions used throughout the paper. Appendix~\ref{apppert} describes the perturbations around the scaling solution $u=u_*$ and $f=f_{1*}x$ important for this work.

\section{Formalism}\label{sectionformalism}
\subsection{The theory}
\noindent We consider an $O(N)$-invariant scalar multiplet $\Phi_a=(\phi_1,\ldots,\phi_N)$ non-minimally coupled to gravity through a Wilsonian action of the form
\begin{equation}\begin{split}
\label{theorysemk}
S_\Lambda[g,\Phi]=\int d^dx\,\sqrt{g}\,\Bigg(- F_\Lambda(\rho)\,R+\frac{1}{2} Z_T(\Lambda)\sum_{a=1}^{N-1}\phi_a\,(-\Box)\phi_a+\frac{1}{2} Z_L(\Lambda)\,\phi_N\,(-\Box)\phi_N+U_\Lambda(\rho)\Bigg),
\end{split}\end{equation}
where $-\Box$ is the Laplace--Beltrami operator acting on the scalar fields and $\rho$ denotes the invariant
\begin{equation}
\rho=\frac{1}{2}\,\Phi_a\Phi_a=\frac{1}{2}\sum_{a=1}^N \phi_a^2 .
\end{equation}
Eq.(\ref{theorysemk}) is based on a derivative expansion \cite{Morris:1994ie,Morris:1997xj,Morris:1994au} in $\partial_\mu\phi$ where only the $O(\partial^2)$ terms are kept. Details on this approximation are given in Sec. \ref{sec:approximations}.

We work in the broken phase and choose the background  direction to lie along the $N$-th component. Accordingly, we decompose the multiplet into one longitudinal (radial) field and $N-1$ transverse (Goldstone) fields,
\begin{equation}
\Phi_a=\phi_N\,n_a+\pi_a,\qquad n_a=\delta_{aN},\qquad n\cdot\pi=0 .
\end{equation}
Equivalently, introducing the longitudinal and transverse projectors (defined with respect to the chosen direction $n_a$),
\begin{equation}\begin{split}
&P^{L}_{ab}=n_a n_b,\qquad P^{T}_{ab}=\delta_{ab}-n_a n_b,
\end{split}\end{equation}
one has $\phi_N=P^{L}_{Nb}\Phi_b$ and $\pi_a=P^{T}_{ab}\Phi_b$ (with $a=1,\ldots,N-1$).

In eq.~(\ref{theorysemk}) we allow for two (scale-dependent but field-independent) wave-function renormalizations, $Z_L(\Lambda)$ for the longitudinal mode and $Z_T(\Lambda)$ for the transverse modes. This corresponds to the LPA$'$-type approximation for the scalar sector. The associated anomalous dimensions are
\begin{equation}
\label{defeta}
\eta_T(\Lambda)=-\frac{\Lambda\,\partial_\Lambda Z_T(\Lambda)}{Z_T(\Lambda)},\qquad
\eta_L(\Lambda)=-\frac{\Lambda\,\partial_\Lambda Z_L(\Lambda)}{Z_L(\Lambda)} ,
\end{equation}
so that the (RG) scaling dimensions of transverse and longitudinal fields differ,
\begin{equation}
[\phi_a]=\frac{d-2+\eta_T}{2},\qquad
[\phi_N]=\frac{d-2+\eta_L}{2}\, .
\end{equation}

\subsection{The proper-time flow equations}
The proper-time flow equations for $F$ and $U$ at $Z=1$ in eq.(\ref{theorysemk}) have been derived in \cite{Bonanno:2025qsc} using the background-field method and the heat-kernel expansion, with linear splitting for the scalar and exponential splitting for the metric, in the physical gauge \cite{Percacci:2015wwa,Ohta:2015fcu}. The flow equations for $Z_L$ and $Z_T$ follow from the same setup; details are given in Appendix~\ref{appZLZT}. It is convenient to introduce the ratio
\begin{equation}
w(\Lambda)=\frac{Z_T(\Lambda)}{Z_L(\Lambda)} ,
\end{equation}
which measures the relative normalization of transverse and longitudinal modes. Its RG evolution is governed by
\begin{equation}
\Lambda\,\partial_\Lambda w=w(\eta_L-\eta_T).
\end{equation}
or equivalently by $\Lambda\,\partial_\Lambda \ln w=\eta_L-\eta_T$.

Introducing the dimensionless variables
\begin{equation}
x=Z_L\,\Lambda^{2-d}\rho,\qquad u_\Lambda(x)=\Lambda^{-d}U_\Lambda(\rho),\qquad f_\Lambda(x)=\Lambda^{2-d}F_\Lambda(\rho),
\end{equation}
we obtain the dimensionless flow equations for $u_\Lambda$ and $f_\Lambda$,
\begin{equation}\begin{split}
\label{adeqfull}
&\dot{u}=-d\,u+\bigl(d-2+\eta_L\bigr)\,x\,u^\prime+\alpha\Bigg(2d(d-3)+4(N-1)\left(1+\frac{u^\prime}{\gamma w}\right)^{\frac{d}{2}-m}+4\,\mathcal{P}^{\,\frac{d}{2}-m}\Bigg),\\
&\dot{f}=(2-d)\,f+\bigl(d-2+\eta_L\bigr)\,x\,f^\prime-\frac{\alpha(d-2m)}{12\gamma}\Bigg(72+6d-2d^2-\\
&-4(N-1)\left(1+\frac{6f^\prime}{w}\right)\left(1+\frac{u^\prime}{\gamma w}\right)^{\frac{d}{2}-m-1}-4\,\mathcal{Q}\,\mathcal{P}^{\,\frac{d}{2}-m-1}\Bigg),
\end{split}\end{equation}
where
\begin{equation}\begin{split}
&\alpha=\frac{\gamma^{d/2}\Gamma\!\left(m-\frac{d}{2}\right)}{4(4\pi)^{d/2}\Gamma(m)},\\
&\mathcal{P}=1+\frac{u^\prime+2xu^{\prime\prime}}{\gamma\left(1+\frac{4(d-1)x(f^\prime)^2}{(d-2)f}\right)},\quad \quad \mathcal{Q}=1+\frac{6\left(f^\prime+\frac{4x(f^\prime)^2}{(d-2)f}+2xf^{\prime\prime}\right)}
{1+\frac{4(d-1)x(f^\prime)^2}{(d-2)f}} .
\end{split}\end{equation}
Here a prime denotes a derivative with respect to $x$, and a dot denotes a derivative with respect to RG time $t=\ln(\Lambda/\Lambda_0)$, with $\Lambda_0$ an arbitrary reference scale.

The anomalous dimensions read
\begin{equation}\begin{split}
\label{etaLetaT}
\eta_L&=\frac{\gamma^{\frac{d}{2}-3}}{3(4\pi)^{d/2}}\frac{\Gamma\!\left(m-\frac{d}{2}+3\right)}{\Gamma(m)}\left(1+\frac{\bar{u}^\prime+2\bar{x}\bar{u}^{\prime\prime}}{\gamma\left(1+\frac{4(d-1)\bar{x}(\bar{f}^\prime)^2}{(d-2)\bar{f}}\right)}\right)^{\frac{d}{2}-(m+3)}
\Biggl[\frac{\bar{x}(\bar{u}^{\prime\prime})^2}{w^2}(N-1)\\
&+\frac{4\bar{x}^2\bar{u}^{\prime\prime\prime}-3\bar{u}^\prime}{4\bar{x}\left(1+\frac{4(d-1)\bar{x}(\bar{f}^\prime)^2}{(d-2)\bar{f}}\right)^2}\left(1+\frac{\bar{u}^\prime+2\bar{x}\bar{u}^{\prime\prime}}{4\bar{x}^2\bar{u}^{\prime\prime\prime}-3\bar{u}^\prime}\left(1+\frac{2\left(1+\frac{4(d-1)\bar{x}^2\bar{f}^\prime}{(d-2)\bar{f}}\left(\frac{(\bar{f}^\prime)^2}{\bar{f}}-2\bar{f}^{\prime\prime}\right)\right)}{1+\frac{4(d-1)\bar{x}(\bar{f}^\prime)^2}{(d-2)\bar{f}}}\right)\right)^2\Biggr],\\
&\eta_T=\frac{\gamma^{\frac{d}{2}-3}}{6(4\pi)^{d/2}\,\bar{x}\,w}\frac{\Gamma\!\left(m-\frac{d}{2}+3\right)}{\Gamma(m)}\left(1+\frac{\bar{u}^\prime}{\gamma w}\right)^{\frac{d}{2}-(m+3)}\left(\frac{\bar{u}^\prime}{w}-\frac{\bar{u}^\prime+2\bar{x}\bar{u}^{\prime\prime}}{1+\frac{4(d-1)\bar{x}(\bar{f}^\prime)^2}{(d-2)\bar{f}}}\right)^2 .
\end{split}\end{equation}
A bar indicates that the corresponding quantity is evaluated on the background configuration used in the derivation of the flow (see Appendix~\ref{appZLZT}).

\subsection{The truncation}
Solving the flow equations Eqs.~(\ref{adeqfull}) for general $u_t(x)$ and $f_t(x)$ is computationally demanding. In the absence of gravity, numerical solutions for the effective potential are typically obtained using pseudo-spectral methods \cite{Borchardt:2016pif,Borchardt:2015rxa} or grid-based approaches \cite{Zorbach:2024rre,Fischbacher:2012ib}. Alternatively, polynomial expansions of the effective potential around its running minimum are widely used:
\begin{equation}\begin{split}
\label{truncpol}
&u_t(x) = u_0(t) + \sum_{n=2}^{\infty} a_n(t)\,(x - x_0(t))^n \\
&f_t(x) = f_0(t) + \sum_{n=1}^{\infty} b_n(t)\,(x - x_0(t))^n
\end{split}\end{equation}
where $x_0(t)$ denotes the running vacuum expectation value (vev), defined by $u_t'(x_0) = 0$. A non-vanishing running minimum $x_0(t)$ at finite RG scales reflects the structure of the RG flow and should not be interpreted as a statement about symmetry breaking, which is determined only in the infrared.

Without loss of generality we choose the vev to point in the $N$ direction, so that only the longitudinal scalar mode acquires an expectation value. Accordingly, we parametrize the $O(N)$ multiplet as
\begin{equation}
\Phi_i=(v+\sigma)\,n_i+\pi_i,\qquad i=1,\ldots,N,\qquad n_i=\delta_{iN},\qquad n\cdot\pi=0,
\end{equation}
and the relation between $v$ and $x_0$ reads $x_0=Z_L\Lambda^{2-d}v^2/2$.

The physical mass of the longitudinal fluctuation is, for general $u$ and $f$,
\begin{equation}
\label{massL}
m_\sigma^2=\Lambda^{2}\, \frac{2x_0\,u_t^{\prime\prime}(x_0)}{1+\frac{4(d-1)x_0\left[f_t^\prime(x_0)\right]^2}{(d-2)\,f_t(x_0)}}\, .
\end{equation}
The $N-1$ transverse fields become Goldstone bosons and remain massless. Since $m_\sigma^2>0$ and we require $u_t''(x_0)>0$ for a positive-definite Hamiltonian, Eq.~(\ref{massL}) implies that $x_0>0$.

\section{Quality and limitations of the approximation scheme}
\label{sec:approximations}
\noindent The results presented in this work rely on a truncation of the Wilsonian action within the functional renormalization group (FRG) framework. As in all FRG studies, the reliability of the conclusions depends on the quality and convergence properties of the approximation scheme. In this section, we summarize the main sources of approximation, discuss their expected impact, and relate them to existing results in the literature.

In the present work, the gravitational sector is treated within a truncation of the form $F(\phi) R$, i.e. we retain only terms linear in the Ricci scalar while allowing for a non-trivial dependence on the scalar fields. This should be distinguished from truncations based on an expansion in curvature invariants, such as $f(R)$ or higher-order operators like $R_{\mu\nu}R^{\mu\nu}$. Such terms are not included in our analysis. The present approximation therefore captures the leading non-minimal coupling between scalar fields and gravity, while neglecting higher-curvature contributions. As a result, quantitative properties of the flow may be affected by these omissions. However, as discussed in the introduction, the existence of interacting ultraviolet fixed points in gravity and gravity--matter systems appears to be a robust feature across different truncations. In this sense, we expect the qualitative mechanism identified in this work to be stable, although a more complete treatment of the gravitational sector would be required for quantitative precision.

Our analysis is based on a derivative expansion truncated at order $\mathcal{O}(\partial^2)$, supplemented by scale-dependent wave-function renormalizations (LPA$'$-type approximation). For purely scalar theories, the convergence of the derivative expansion has been extensively investigated (see e.g. \cite{Morris:2000hm,Canet:2002gs,Litim:2001up,Litim:2001dt,Canet:2003qd,Litim:2010tt,DePolsi:2020pjk}). These studies show that universal quantities such as critical exponents converge rapidly as higher-derivative operators are included. In particular, already at $\mathcal{O}(\partial^2)$ one typically obtains percent-level accuracy for critical exponents in three dimensions. In the presence of gravity, a comparable level of quantitative control is not yet available. Nevertheless, the derivative expansion has been widely employed in asymptotically safe gravity and gravity–matter systems (see e.g. \cite{Percacci:2003jz,Narain:2009fy, Narain:2009gb,Percacci:2015wwa, Labus:2015ska,Bonanno:2025qsc}), where it consistently leads to the emergence of interacting ultraviolet fixed points. While the convergence of the expansion is less well established than in scalar theories, generally it provides a well-defined and systematically improvable framework, expected to capture the dominant contributions to the flow at the present level of truncation.

We approximate the functional dependence of the scalar potential and the non-minimal coupling by a polynomial expansion around the running minimum of the potential. In scalar theories, such an expansion leads to a rapidly convergent truncation scheme \cite{Morris:1994ki,Aoki:1998um}. In particular, these expansions exhibit improved convergence properties compared to expansions around the origin \cite{WEYRAUCH200621}. However, they remain local approximations and may fail to capture global features of the potential, such as non-analyticities or metastable regions \cite{Dupuis:2020fhh,Berges:2000ew}. In the present work, our focus is on the existence of UV-complete trajectories and on the mechanism that removes the Landau-pole singularity. These features are governed by the behavior of the flow in a finite region of field space and are therefore expected to be less sensitive to the global structure of the potential. Nevertheless, quantitative predictions in the infrared may be affected by the truncation.

Functional renormalization group flows depend on the choice of regulator, gauge, and field parametrization. In gravity, this dependence is known to affect quantitative results, such as the number of relevant directions and the values of critical exponents \cite{Falkenberg:1996bq,Souma:2000vs,Barra:2019rhz,Bonanno:2025tfj,Ohta:2016npm,Ohta:2016jvw}. At the same time, certain qualitative features—most notably the existence of interacting ultraviolet fixed points in gravity and gravity--matter systems—are typically found to persist across different setups. In this work, on the one hand we partially assess scheme dependence by varying the cutoff shape parameter and considering different families of regulators. As shown in Sec.~\ref{sectioncriticalline}, the existence and qualitative properties of the UV-complete trajectories are only mildly affected by these variations. On the other hand, we do not perform a systematic comparison with alternative choices of parametrization or different gauge fixings. Throughout this work, we adopt the exponential parametrization of the metric together with the physical gauge. As a consequence, we do not attempt to quantify how sensitive specific results — such as fixed-point values, critical exponents, or the precise location of the critical surface — are to these choices. The results presented here should therefore be interpreted within the specific scheme defined by the chosen parametrization and gauge. A more complete assessment of parametrization and gauge dependence would require a systematic comparison across different choices, which we leave for future work.

Evidence from scalar theories and asymptotically safe gravity indicates that such approximations can capture the qualitative structure of the RG flow, in particular the presence of interacting fixed points and the dimensionality of the critical surface. In this light, the results obtained in this work are expected to be qualitatively robust. At the same time, quantitative predictions—especially in the infrared regime—should be interpreted with appropriate caution. They may be affected by higher-order operators, improved truncations, and a more systematic assessment of parametrization and gauge dependence, which we leave for future work.

\section{The properties of the RG flow}\label{sectionbeta}
\noindent In this section we discuss how the presence of gravity in the flow of the running couplings in eq.(\ref{truncpol}) is able to produce a mechanism that induces a suppression of Landau poles, leading to UV complete trajectories.

\subsection{The scaling solutions}\label{secexscalsol}
Within the truncation of the Wilsonian action adopted in Eq.~(\ref{theorysemk}) with $Z=1$, it was shown in \cite{Bonanno:2025qsc} that $u_* = u_0$ and $f_*(x) = f_0 + f_1 x$, both for $f_0 \neq 0$ and $f_0 = 0$, define two independent classes of exact scaling solutions of Eqs.~(\ref{adeqfull}). These scaling solutions persist upon including a wavefunction renormalization within the same truncation. Closely related scaling solutions in scalar--tensor systems with the same truncation of the effective action, gauge and parametrization choice have also been reported within the Wetterich formalism, see e.g.~\cite{Percacci:2015wwa,Labus:2015ska}.

In $d=4$, the scaling constant potential and $w_*$ are given by
\begin{equation}
u_* = \gamma^2 \frac{2+N}{64\pi^2} \frac{\Gamma(m-2)}{\Gamma(m)}, \qquad w_* = a \, ,
\end{equation}
where $w_*$ is a free parameter, denoted by $a$. This reflects the fact that, for a constant scaling potential, one has
\begin{equation}
\eta_{L*} = 0, \qquad \eta_{T*} = 0 \, ,
\end{equation}
so that the ratio $w = Z_T/Z_L$ is not fixed by the flow.

The first class consists of two interacting Reuter-type fixed points in the presence of matter:
\begin{equation}
\label{scalex1}
f_*(x) = \gamma \frac{16 - N}{192\pi^2 (m-1)} \, ,
\end{equation}
and
\begin{equation}
\label{scalex2}
f_*(x) = \gamma \frac{2(1 - N) + (14 - N) w_*}{192 (m-1)\pi^2 w_*} + \frac{1}{3} x \, .
\end{equation}
These expressions coincide with the scaling solutions found in \cite{Bonanno:2025qsc} for $f_1 = 0$ and $f_1 \neq 0$, respectively, and are here extended to the case $Z \neq 1$ within the same truncation.

The second class consists of two scaling solutions given by
\begin{equation}
\label{scalex34}
f_*^{(\pm)}(x) = -\frac{1}{12} \left( 1 + \frac{N - 13 \pm \sqrt{(N-13)^2 - \frac{2}{w_*}(N-19)(N-1) + \frac{1}{w_*^2}(N-1)^2}}{N-1} w_* \right) x \, .
\end{equation}
As in \cite{Bonanno:2025qsc}, the branch with the minus sign yields $f_{1*} < 0$ for $N < 16$ throughout the parameter region (see Fig.~\ref{plotf1fp2}), whereas the branch with the plus sign gives $f_{1*} > 0$ (see Fig.~\ref{plotf1fp3}). Moreover, only the minus-sign branch admits a regular limit as $N \to 1$.

The constant term in $f_*(x)$ for the second class coincides with that of the Reuter fixed point in the first class. Furthermore, as in \cite{Bonanno:2025qsc}, the value of $f_{1*}$ does not depend on $m$ and, consequently, is also independent of $\gamma$.

The perturbations of the first class of scaling solutions have already been analyzed in \cite{Bonanno:2025qsc}. The perturbations of $f_*^{(+)}$ in the second class, which are particularly relevant for the present work, are discussed in Appendix~\ref{apppert}.

\subsection{The truncation}
Within the present work, we truncate Eq.~(\ref{truncpol}) by retaining only the lowest-order terms in the expansion of the effective potential around the running minimum. In this way, we obtain:
\begin{equation}\begin{split}
\label{trunclow}
&u_t(x)=u_0(t)+\lambda(t)\bigl(x-x_0(t)\bigr)^2\\
&f_t(x)=\tilde{m}_p^2(t)+f_1(t)\bigl(x-x_0(t)\bigr)
\end{split}\end{equation}
where we set $a_2(t)=\lambda(t)$, $f_0(t)=\tilde{m}_p^2(t)$ and $b_1(t)=f_1(t)$. Here $\tilde{m}_p(t)$ is the dimensionless Plank mass. In the absence of $f_1$ the inverse of $\tilde{m}_p^2$ corresponds to the dimensionless Newton coupling. We therefore set 
\begin{equation}
\tilde{m}_p^2(t)=\frac{1}{g(t)}.
\end{equation}
The truncation in Eq.~(\ref{trunclow}) captures the leading features of the RG flow relevant for the analysis performed in this work. In particular, the existence and qualitative structure of the UV-complete trajectories discussed in the following sections are not sensitive to the inclusion of higher-order operators within the present truncation scheme. However, away from the UV regime, the quantitative reliability of the results is expected to decrease, and the present truncation should be regarded as a first approximation. A more refined description of the flow in these regions would require the inclusion of higher-order terms in the derivative and potential expansions of the effective action.

The truncation for $f_t(x)$ can be written equivalently as a truncation around the origin. Setting 
\begin{equation}
\label{relmp}
\tilde M_{p}(t)=\tilde{m}_p^2(t)-f_1(t)x_0(t)
\end{equation}
we get
\begin{equation}
\label{truncor1}
f_t(x)=\tilde M_p(t)+f_1(t) x
\end{equation}
this equivalence is useful to analyze the fixed-point solutions and their properties because it allows to relate the truncation eq.(\ref{trunclow}) to the exact scaling solutions eq.(\ref{scalex2}) and (\ref{scalex34}).

Using eq.(\ref{trunclow}), the dimensionless mass corresponding to Eq.~(\ref{massL}) becomes
\begin{equation}
\tilde{m}_\sigma^2=\frac{4x_0\lambda}{1+\frac{4(d-1)x_0 f_1^2}{(d-2)\,\widetilde{m}_p^2}}
=\frac{4x_0\lambda}{1+\frac{4(d-1)g x_0 f_1^2}{d-2}}\, .
\end{equation}
In the absence of gravitational interactions, or for $f_1=0$, this reduces to the standard result.

\subsection{The conditions on the flow of \texorpdfstring{$\lambda$}{N} for a regular running}
The beta functions for $u_0$, $x_0$, $\lambda$, $g$, $f_1$, and $w$ are derived in Appendix~\ref{appbeta}. For the present discussion we focus on the flow of $\lambda$. Generally, the flow is composed by two contributions:
\begin{equation}
\label{betafun}
\dot{\lambda}(t)=\beta_g(t)+\beta_m(t),
\end{equation}
where $\beta_g$ and $\beta_m$ are the gravitational and matter contribution to the flow respectively. In particular, the two contributions can be written as
\begin{equation}
\beta_g(t)=-A(t)\lambda(t),\quad \quad \beta_m(t)=\eta_L(t)\lambda(t)+b(t)\lambda^2(t)
\end{equation}
where in $d=4$ the two coefficients $A$ and $b$ are given by
\begin{equation}\begin{split}
&A(t)=-\frac{9\gamma f_1^2g\left(1-f_1gx_0\right)\left(1+2f_1gx_0\left(f_1-\frac{1}{3}\right)\right)}{4\pi^2(m-1)\left(1+6f_1^2gx_0\right)^3\left(1+\frac{4\lambda x_0}{\gamma\left(1+6f_1^2gx_0\right)}\right)^m}
\end{split}\end{equation}
and
\begin{equation}\begin{split}
&b(t)=\frac{1}{8\pi^2(m-1)}\left(\frac{(m-1)(N-1)}{w^2}+\frac{3C(t)}{\left(1+6f_1^2gx_0\right)^4\left(1+\frac{4\lambda x_0}{\gamma\left(1+6f_1^2gx_0\right)}\right)^m}\right),\\[1mm]
&C(t)=3(m-1)+12g(1+m)f_1^2x_0+4f_1^3g^2x_0^2\left(6m-16+3(3+m)f_1\right)\\
&\hspace{1.0cm}+16f_1^4g^3x_0^3\left(1+3(m-2)f_1\right)+48f_1^6g^4x_0^4(m-1).
\end{split}\end{equation}
The expression for $\eta_L(t)$ is given in Eq.~(\ref{betaetaL}). This contributes with a term $\eta_L\sim\lambda^2$ so the piece $\lambda\eta_L$ is actually of the third order in $\lambda$. It is convenient to write the matter contribution as
\begin{equation}
\beta_m(t)=\eta_L(t)\lambda(t)+b(t)\lambda^2(t)=(D(t)\lambda(t)+b(t))\lambda^2(t)\equiv B(t) \lambda^2(t)
\end{equation}
where we used $\eta_L(t)=D(t)\lambda^2(t)$. The expression for $D(t)$ can be read from eq.~(\ref{betaetaL}).

The presence of a non-vanishing non-minimal parameter $f_1$ couples $\lambda$ to the gravitational sector through the linear term proportional to $A(t)$. For $f_1=0$ one recovers the purely scalar structure, in which gravity does not affect the familiar one-loop form and the coefficient $b(t)$ of $\lambda^2$ in $\beta_m(t)$ is positive. In the present truncation and for the parameter range considered in this work, the coefficient $b(t)$ remains positive for $f_1\neq 0$ too. Furthermore, it turns out that $D(t)\ge0$ always and accordingly $B(t)>0$, therefore the matter contribution to the flow of $\lambda$ always drives the solution to a singularity in this case.

The linear term in $\beta_g$ can be interpreted as providing an effective scaling contribution for $\lambda$. With the convention in Eq.~(\ref{betafun}), antiscreening corresponds to $A(t)>0$, since then the term $\beta_g(t)=-A(t)\lambda$ drives $\lambda$ towards smaller values as the RG scale increases. Since $x_0>0$ and $g>0$, the gravitational contribution to $A(t)$ can change sign only for $f_1>0$. If $A(t)$ becomes positive and dominates the dynamics from some scale $\bar t$ up to $t\to\infty$, the singularity is avoided and $\lambda$ decreases towards the ultraviolet.

More generally, in many gravity--matter systems gravitational fluctuations can induce antiscreening contributions that weaken matter self-interactions at high scales \cite{Shaposhnikov:2009pv,Wetterich:2016uxm,Hamada:2017rvn,Eichhorn:2017ylw,Eichhorn:2018whv,Pawlowski:2018ixd,Wetterich:2019zdo,Pastor-Gutierrez:2022nki,Garces:2025rgn}. However, not all RG trajectories are expected to approach a UV fixed point. In our case the relevant question is under which conditions the gravitational fluctuations controlled by $f_1$ and $g$ overcome the matter fluctuations, i.e. when the condition $|\beta_g(t)|>\beta_m(t)$ is reached. From Eq.~(\ref{betafun}) this requires that the linear term becomes sufficiently important compared to the quadratic and cubic terms in $B(t)$, so that the flow can enter a regime where $\lambda$ is driven away from the singularity and towards a scale-invariant trajectory. 

In such a regime, one expects $\lambda$ to approach an interacting scaling value $\lambda_I$, defined implicitly by the condition
\begin{equation}
\lambda_I=\frac{A_I}{B_I} \,,
\end{equation}
with $A_I>0$. Since both $A_I$ and $B_I$ depend on $\lambda$, this relation should be understood as an implicit equation determining $\lambda_I$. Equivalently, it corresponds to a zero of the beta function, $\beta_\lambda(\lambda_I)=0$, i.e. to a stationary point of the flow of $\lambda$ within the present truncation, rather than to a fixed point of the full RG system. 

Once this regime is reached, the condition $|A|>B\lambda$ signals that the gravitational contribution $\beta_g(t)=-A(t)\lambda(t)$ dominates over the matter contribution $\beta_m(t)=B(t)\lambda^2(t)$, thereby ensuring a further decrease of $\lambda$ towards the UV.

Since in the low-energy regime one typically has $A<0$, this mechanism requires a crossover at which the linear term changes sign. Motivated by the explicit factor $(1-f_1gx_0)$ appearing in $A(t)$, one may estimate the corresponding condition as
\begin{equation}
\label{critval}
f_1^{\mathrm{crit}}(t)=\frac{1}{g(t)\,x_0(t)}\, .
\end{equation}
For $f_1(t)\ge f_1^{\mathrm{crit}}(t)$ the linear term can become negative so that $|\beta_g(t)|>\beta_m(t)$, allowing for a positive scaling value $\lambda_I>0$ and, for sufficiently large $t$, a decay of $\lambda$ towards the ultraviolet. Given a numerical solution, Eq.~(\ref{critval}) also provides an estimate for the scale $\bar t$ at which the crossover occurs. The condition $f_1(t)\ge f_1^{\mathrm{crit}}(t)$ together with $|A|>B\lambda$ for $t>\bar t$ implies that, instead of growing logarithmically towards the UV, $\lambda$ decreases according to some inverse power law in $t$ and becomes asymptotically free as $t\to\infty$.

\section{Fixed-point and perturbations}\label{sectionfixedpoint}
\subsection{Fixed points}
\noindent If the conditions discussed in Sec.~\ref{sectionbeta} are satisfied, $\lambda$ does not run into a singularity. In fact, the absence of singularities at large RG time requires that the flow approaches a fixed-point regime as $t\to\infty$. The fixed points therefore provide the main tool to understand the global RG dynamics. 

Using the beta functions reported in Appendix~\ref{appbeta}, we find that all fixed points share
\begin{equation}
u_{0\ast}=\gamma^2\frac{2+N}{64\pi^2}\frac{\Gamma(m-2)}{\Gamma(m)},\qquad \lambda_\ast=0,\qquad w_\ast=a,
\end{equation}
the values of $u_*$ and $w_*$ are the same of the scaling solutions of section \ref{secexscalsol}.

\begin{figure}[p]
     \centering
     \subfigure[]{
         \centering
         \includegraphics[width=0.45\textwidth]{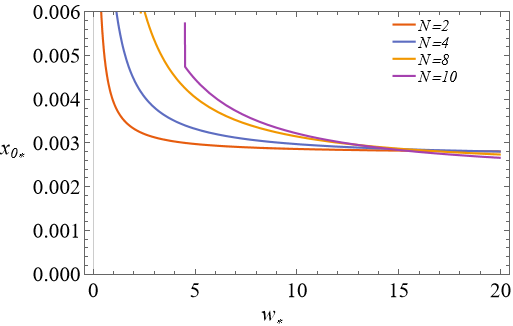}
         \label{plotx0fp1}
     }
    \hspace{0.8em}
     \subfigure[]{
         \centering
         \includegraphics[width=0.45\textwidth]{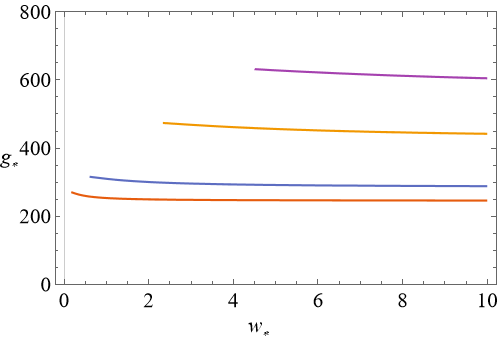}
         \label{plotgfp1}
     }
     \caption{Fixed-point values for the family with $f_{1\ast}=1/3$, shown for $m=d/2+1$ and $\gamma=1$ as functions of $w_\ast=a$ for representative values of $N$. (a) $x_{0\ast}$ versus $w_\ast$. (b) $g_\ast$ versus $w_\ast$.}

\vspace{0.5cm}

     \centering
     \subfigure[]{
         \centering
         \includegraphics[width=0.45\textwidth]{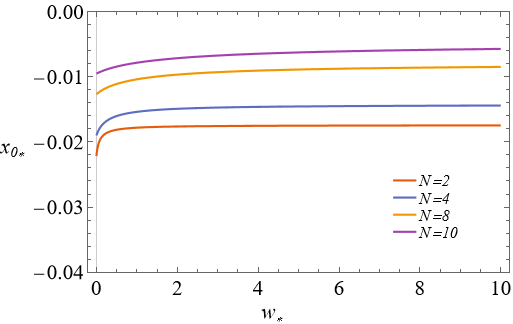}
         \label{plotx0fp2}
     }
    \hspace{0.8em}
     \subfigure[]{
         \centering
         \includegraphics[width=0.45\textwidth]{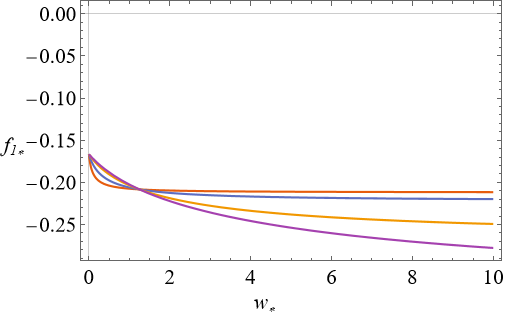}
         \label{plotf1fp2}
     }
     \caption{Fixed-point values for the family defined by Eq.~(\ref{FPbello}) with the minus sign, shown for $m=d/2+1$ and $\gamma=1$ as functions of $w_\ast=a$ for representative values of $N$. (a) $x_{0\ast}$ versus $w_\ast$. (b) $f_{1\ast}$ versus $w_\ast$.}

\vspace{0.5cm}

       \centering
     \subfigure[]{
         \centering
         \includegraphics[width=0.45\textwidth]{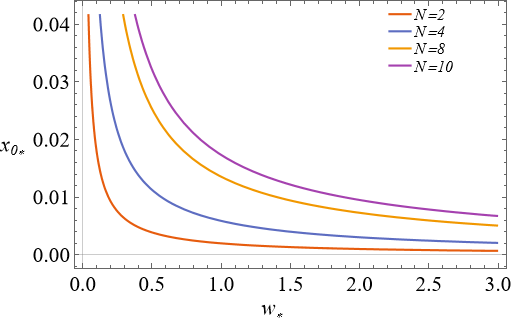}
         \label{plotx0fp3}
     }
    \hspace{0.8em}
     \subfigure[]{
         \centering
         \includegraphics[width=0.45\textwidth]{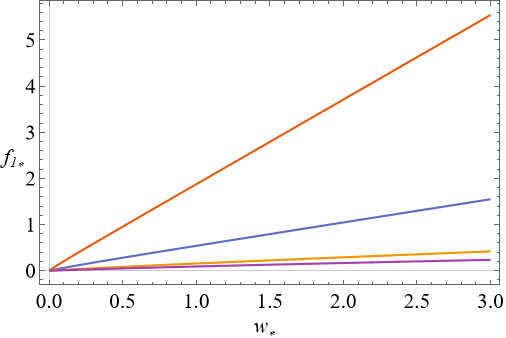}
         \label{plotf1fp3}
     }
     \caption{Fixed-point values for the family defined by Eq.~(\ref{FPbello}) with the plus sign, shown for $m=d/2+1$ and $\gamma=1$ as functions of $w_\ast=a$ for representative values of $N$. (a) $x_{0\ast}$ versus $w_\ast$. (b) $f_{1\ast}$ versus $w_\ast$.}
\end{figure}

Concerning the remaining couplings $x_{0*}$, $g_*$, and $f_{1*}$, we find five fixed points, which can be organized according to the classification adopted for the two classes of scaling solutions in Eqs.~(\ref{scalex1}), (\ref{scalex2}), and (\ref{scalex34}). The first class includes the Gaussian fixed point,
\begin{equation}
x_{0\ast}=\gamma\frac{N-1+3a}{32\pi^2(m-1)a},\qquad g_\ast=0,\qquad f_{1\ast}=b,
\end{equation}
where also $f_{1\ast}$ is free (we denote it by $b$), and the two interacting Reuter-type fixed points in the presence of matter,
\begin{equation}
\label{FP2}
x_{0\ast}=\gamma\frac{N-1+3a}{32\pi^2(m-1) a},\qquad g_\ast=\frac{192\pi^2(m-1)}{(16-N)\gamma},\qquad f_{1\ast}=0,
\end{equation}
and
\begin{equation}
\label{FP3}
x_{0\ast}=h(N,a,m),\qquad g_\ast=l(N,a,m),\qquad f_{1\ast}=\frac{1}{3}.
\end{equation}
At the Reuter fixed point with $f_{1*} = 0$, corresponding to Eq.~(\ref{scalex1}), the value of $g_*$ is simply given by the inverse of $f_{0*}$. By contrast, for the fixed point with $f_{1*} = 1/3$, associated with Eq.~(\ref{scalex2}), $g_*$ is no longer given by $g_* = 1/f_{0*}$, but instead becomes a non-trivial function of $N$, $a$, and $m$. In this case, a non-trivial value of $x_{0*}$ is also present. These quantities are shown in Figs.~\ref{plotx0fp1} and \ref{plotgfp1} for $m = 3$ and $\gamma = 1$, as functions of $w_* = a$ and for representative values of $N$.

The second class, related to eq.(\ref{scalex34}), consists of two fixed points,
\begin{equation}\begin{split}
\label{FPbello}
&x_{0\ast}=\frac{\gamma}{64\pi^2(m-1)}\left(N-13+\frac{N-1}{a}\pm \sqrt{A}\right),\qquad
g_\ast=\frac{192\pi^2(m-1)}{(16-N)\gamma},\\
&f_{1\ast}=-\frac{1}{12}\left(1+\frac{N-13\pm\sqrt{(N-13)^2-\frac{2}{a}(N-19)(N-1)+\frac{1}{a^2}(N-1)^2}}{N-1}a\right),
\end{split}\end{equation}
The values of $f_{1*}$ coincide with those in Eq.~(\ref{scalex34}) and, in particular, satisfy the relation
\begin{equation}
\frac{1}{g_*} = f_{1*} x_{0*} \, .
\end{equation}
Accordingly, using Eqs.~(\ref{relmp}) and (\ref{truncor1}), one recovers $f = f_{1*} x$ and $u=u_*$. Therefore, these fixed points represent a generalization of the scaling solutions in Eq.~(\ref{scalex34}) when a running potential is present. The branch with the minus sign yields $x_{0*} < 0$ for $N < 16$ throughout the parameter region (see Fig.~\ref{plotx0fp2}). By contrast, the branch with the plus sign gives $x_{0*} > 0$ (see Fig.~\ref{plotx0fp3}). As in Eq.~(\ref{scalex34}), only the minus-sign branch admits a regular limit as $N \to 1$.

Extending the polynomial truncation does not generate new non-trivial values for the fixed-point couplings:
\begin{equation}
u_{*n} = 0, \qquad f_{*n} = 0, \qquad n > 2 \, .
\end{equation}
Thus, the higher-order couplings do not alter the previous conclusions. In particular, for all fixed points, the value of $f_{1*}$ remains independent of $m$ and $\gamma$. Moreover, apart from the special case in which $f_{1*} = 1/3$, the fixed-point value of $g_*$ is the same in all cases and is given by
\[
g_* = \frac{192\pi^2 (m-1)}{\gamma (16 - N)} \, .
\]
It is therefore defined only for $N \neq 16$ and becomes negative for $N > 16$.

Since the inclusion of running couplings does not modify the scaling solutions in Eqs.~(\ref{scalex1}), (\ref{scalex2}), and (\ref{scalex34}), these still provide, within the truncation in Eq.~(\ref{theorysemk}), exact scaling regimes that can be reached from fully non-trivial running potentials $u_t(x)$ and $f_t(x)$. By contrast, the running couplings modify the structure of Eq.~(\ref{scalex2}), such that the corresponding scaling regime becomes an isolated point in the space of solutions and cannot be reached from any non-trivial running potential $u_t(x)$ and $f_t(x)$.

The family structure of the fixed point does not depend on $\gamma$. The fact that qualitative fixed-point physics does not depend on $\gamma$ suggests that the corresponding numerical RG trajectories should display the same $\gamma$-independence. 

Finally, the appearance of exact scaling solutions $u=u_\ast$ and $f=f_{0\ast}+f_{1\ast}x$ and the associated fixed-point structure is not specific to the proper-time flow, see e.g.~\cite{Labus:2015ska,Percacci:2015wwa}.

\subsection{Critical exponents}\label{sectioncriticalexp}
Replacing the fixed-point values into the critical condition in Eq.~(\ref{critval}), we find that only the fixed points in the second class can satisfy it. Since we require $f_1>0$, only the solution with the plus sign in Eq.~(\ref{FPbello}) remains. If this fixed point governs the ultraviolet (UV) regime, it can yield asymptotically free solutions for $\lambda$. Whether this happens depends on the spectrum of critical exponents.

\begin{figure}[t]
     \centering
     \subfigure[]{
         \centering
         \includegraphics[width=0.45\textwidth]{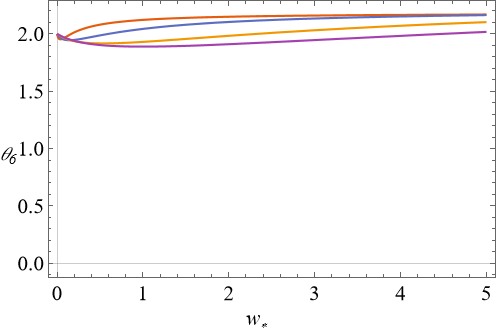}
         \label{plottheta2}
     }
    \hspace{0.8em}
     \subfigure[]{
         \centering
         \includegraphics[width=0.45\textwidth]{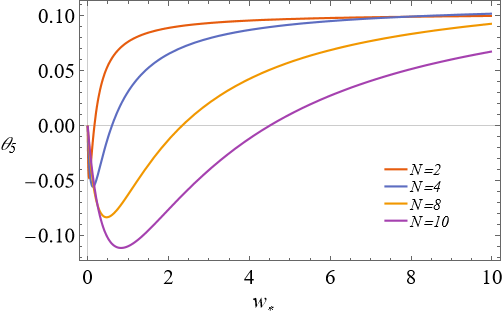}
         \label{plottheta1}
     }
     \caption{Nontrivial critical exponents $\theta_5$ and $\theta_6$ associated with perturbations of the fixed point (\ref{FPbello}) (plus-sign branch), shown as functions of $w_\ast$ for representative values of $N$. (a) $\theta_5(w_\ast)$. (b) $\theta_6(w_\ast)$. While $\theta_5$ remains positive, $\theta_6$ changes sign at $w_\ast=\bar w_\ast(N)$.}
\end{figure}

We linearize the beta functions around the fixed point $h_{i\ast}$ by perturbing the running couplings according to
\begin{equation}
h_i(t)=h_{i\ast}+e^{-t\theta}\,\delta h_i,
\end{equation}
where $h_i\in\{u_0,x_0,\lambda,g,f_1,w\}$ and $\theta$ denotes a critical exponent. For the fixed point eq.~(\ref{FPbello}) with the plus sign we find a set of critical exponents that contains the canonical values $\theta=0$, $\theta=2$, and $\theta=4$, as well as two additional non-trivial exponents, denoted by $\theta_5$ and $\theta_6$, which depend on $w_\ast$. Figures~\ref{plottheta2} and \ref{plottheta1} show $\theta_5$ and $\theta_6$ as functions of $w_\ast$ for representative values of $N$. The exponent $\theta_5$ is always positive and typically of order $\theta_5\simeq 2$. By contrast, $\theta_6$ crosses zero at a value $w_\ast=\bar w_\ast(N)$. For instance, for $N=4$ one finds $\theta_6>0$ for $w_\ast\gtrsim 0.6$. In the following we focus on the parameter region $w_\ast>\bar w_\ast(N)$, for which both non-trivial exponents are positive and the fixed point is UV-attractive in all directions relevant for the UV completion mechanism discussed in Sec.~\ref{sectionbeta}.

Inspecting the corresponding eigenvectors, the exponent $\theta=4$ is associated with the $u_0$ direction, whereas the remaining eigendirections are non-trivial superpositions in the $(x_0,\lambda,g,f_1,w)$ subspace. Nevertheless, their projections show that the most relevant deformation has a dominant component along the Newton coupling $g$. This indicates that the UV scaling regime is largely controlled by the gravitational sector. The marginal and near-marginal eigendirections are also dominated by the gravitational sector but with subleading admixtures of $\lambda$ and $f_1$. In particular, the fixed point breaks the classical marginality of $f_1$ and renders it relevant:
\begin{equation}
f_1(t\to\infty)=f_{1*}^{(+)}+a_1 e^{-\theta_5t}+a_2 e^{-\theta_6t}
\end{equation}
where $a_1$ and $a_2$ are two free coefficients. At linear order, $\lambda$ is marginal in the fixed-point regime; beyond linear order it acquires logarithmic corrections, corresponding to a marginally (ir)relevant behavior: 
\begin{equation}
\lambda(t\to\infty)=\frac{b_1}{t}+\frac{b_2}{t^2}+O\left(\frac{1}{t^3}\right)
\end{equation}
where $b_1$, and $b_2$ are complicated numbers depending on $N$ and $m$. 

In the parameter region $w_\ast>\bar w_\ast(N)$, the fixed point is UV-attractive in the sense required for the UV completion mechanism described above. Since the existence of the plus-sign fixed point requires $N>1$, the singularity can be removed only for an $O(N)$ multiplet with at least two scalar fields.

From the beta functions in Appendix~\ref{appbeta}, the anomalous dimension $\eta_L$ also contributes to the breaking of classical marginality. However, our qualitative conclusions do not hinge on these contributions: setting $Z_L=Z_T=1$ leaves the fixed-point mechanism and the associated relevance properties unchanged. The breaking of marginality is thus a genuine feature of the fixed point.

Extending the polynomial truncation in Eq.~(\ref{truncpol}) does not modify the physical structure of the UV critical manifold. Its only effect is to introduce additional irrelevant directions with $\theta < 0$, while the non-trivial critical exponents $\theta_5$ and $\theta_6$ approach their exact values in the limit $n \to \infty$. The general properties of the UV critical manifold, derived from perturbations of the scaling solutions in Eq.~(\ref{scalex34}), as well as their relation to polynomial truncations around the minimum, are discussed in more detail in Appendix~\ref{apppert}. There, we also present the general formula for the critical exponents and show explicitly that they are independent of the regulator parameters $m$ and $\gamma$.

The RG flow on the UV critical surface spanned by the UV fixed point yields definite predictions for the infrared (IR) values of couplings at a reference scale $t=t_0$. In practice, the projection of the UV critical surface onto the IR hyperplane spanned by $(x_0(t_0),\lambda(t_0),g(t_0),f_1(t_0))$ partitions the space of IR initial conditions into two regions, which we denote by $M_1$ and $M_2$. Initial conditions in $M_1$ correspond to trajectories that remain regular for all scales and are attracted to the UV fixed point: for these trajectories gravitational fluctuations overcome matter fluctuations around the Planck scale and the conditions in Sec.~\ref{sectionbeta} are satisfied, yielding UV-complete theories. By contrast, initial conditions in $M_2$ lie outside the basin of attraction of the UV fixed point: matter fluctuations continue to dominate above the Planck scale, the conditions in Eq.~(\ref{critval}) and $|A|>B\lambda$ are not met, and the flow develops a Landau-pole-like singularity in the UV. Accordingly, points in $M_2$ correspond to trajectories that cannot be connected to the UV fixed point.

Since matter fluctuations dominate in the deep IR, following UV-complete trajectories towards $t \to -\infty$ typically brings them close to the Gaussian regime, where the couplings run according to their canonical scaling. However, reaching the Gaussian fixed point itself requires a tuning of the IR-relevant directions. In particular, the running of the minimum generally includes contributions such as $x_0(t)\sim e^{-2t}$, reflecting the presence of relevant directions around the Gaussian fixed point. As a consequence, there exists a continuous family of RG trajectories approaching the Gaussian regime, while only a subset of them reaches the Gaussian fixed point exactly.

\section{Numerical analysis to determine the critical line}\label{sectioncriticalline}
\noindent Determining the manifolds $M_1$ and $M_2$ requires a numerical analysis. As in the fixed-point discussion, we use $\gamma=1$ for illustration. We solve the coupled beta functions for the running couplings $(u_0,x_0,\lambda,g,f_1,w)$ given in Appendix~\ref{appbeta}, at fixed $N$ and $m$. As boundary conditions we specify the couplings at $t=0$, corresponding to $\Lambda=\Lambda_0$, i.e.\ the reference scale at which the IR parameters are defined, and integrate towards increasing RG time (towards the UV). 

Trajectories that develop a singularity (belonging to $M_2$) terminate at a finite RG time $t=t_{\mathrm{pole}}$, defined operationally as the smallest value of $t$ at which the flow becomes singular (e.g.\ by a divergence of a coupling or by a vanishing denominator in the beta functions). This induces a separating surface in the space of initial conditions, which we may schematically represent as
\begin{equation}
0=S\left(t_{\mathrm{pole}};x_0(0),\lambda(0),g(0),f_1(0),w(0)\right),
\end{equation}
where $S$ encodes the condition that a singularity is reached at $t_{\mathrm{pole}}$. In the purely scalar limit $f_1(0)=0$ one recovers the standard one-loop behavior $\lambda^{-1}(t)=\lambda^{-1}(0)+\beta_0 t$, which implies a divergence at a finite $t_{\mathrm{pole}}$ for $\lambda(0)>0$ and $S$ reduces to $S=1+\beta_0 \lambda(0)t_{pole}=0$ with $\beta_0<0$. For $f_1(0)\neq 0$, the existence of the UV-attractive fixed point discussed in Sec.~\ref{sectionfixedpoint} implies that there are initial conditions for which $\lambda(t)$ remains regular for all $t$ and flows to $\lambda_\ast=0$. Accordingly, the set of initial conditions that develop a singularity forms a bounded region in the coupling space, separated from the regular region by a critical surface.

To visualize this critical surface, we fix $(x_0(0),g(0),w(0))$ and explore the plane of initial conditions $(f_1(0),\lambda(0))$. In this two-dimensional slice, the separating surface reduces to a critical line, which we parametrize as
\begin{equation}
\lambda(0)=h\!\left(f_1(0)\right).
\end{equation}
We then repeat this analysis by varying $x_0(0)$, $g(0)$, and $w(0)$ one at a time, and subsequently scanning also over
$m$ and $N$.

\begin{figure}[p]
     \centering
     \subfigure[]{
         \centering
         \includegraphics[width=0.45\textwidth]{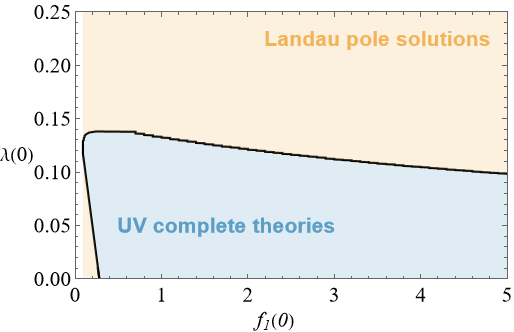}
         \label{plotex}
     }
    \hspace{0.8em}
     \subfigure[]{
         \centering
         \includegraphics[width=0.45\textwidth]{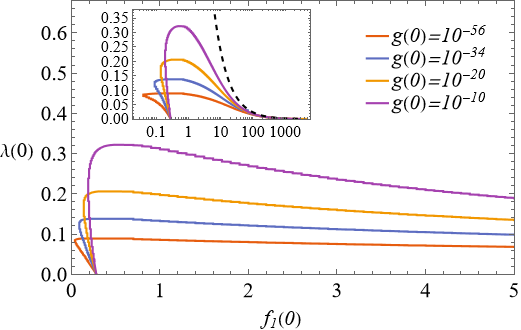}
         \label{plotg}
     }
     \caption{Projection of the IR coupling plane $(f_1(0),\lambda(0))$ at fixed $(x_0(0),g(0),w(0))$, illustrating the separation between UV-complete and singular trajectories. (a) Example for $g(0)=10^{-34}$, $x_0(0)=0.1$, $w(0)=1$ and $N=4$: the region corresponding to UV-complete trajectories (projection of $M_1$) is bounded by the critical line $\lambda(0)=h(f_1(0))$ (black). Outside this region, trajectories develop a singularity (projection of $M_2$). (b) The critical line $\lambda(0)=h(f_1(0))$ for several values of $g(0)$ at fixed $x_0(0)$ and $w(0)$. In the inset the full range up to $f_1(0)\to\infty$. The black dashed line is the asymptotic limit eq.(\ref{critlineasint}).}

\vspace{0.5cm}

     \centering
     \subfigure[]{
         \centering
         \includegraphics[width=0.45\textwidth]{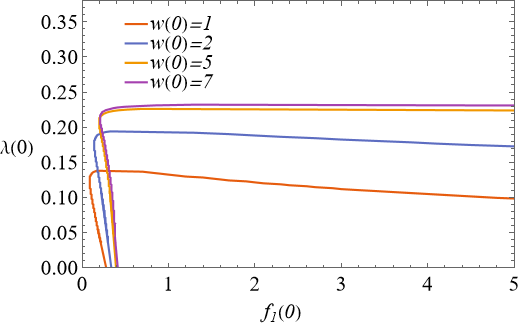}
         \label{plotw}
     }
    \hspace{0.8em}
     \subfigure[]{
         \centering
         \includegraphics[width=0.45\textwidth]{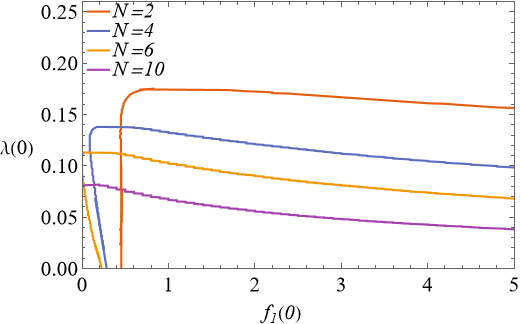}
         \label{plotN}
     }
     \caption{Dependence of the critical line $\lambda(0)=h(f_1(0))$ on IR parameters. (a) Variation with $w(0)$ at fixed $N=4$ and fixed $(g(0),x_0(0))$. (b) Variation with $N$ at fixed $(g(0),x_0(0),w(0))$.}

\vspace{0.5cm}

     \centering
     \subfigure[]{
         \centering
         \includegraphics[width=0.42\textwidth]{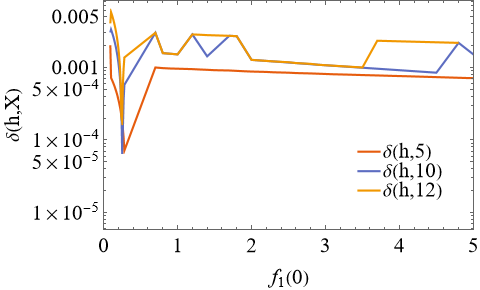}
         \label{diffmpl}
     }
    \hspace{0.8em}
     \subfigure[]{
         \centering
         \includegraphics[width=0.42\textwidth]{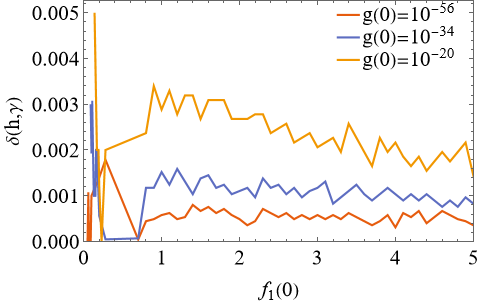}
         \label{diffgamma}
     }
     \caption{Robustness of the critical line under changes of the cutoff scheme. (a) $\delta(h,X)=|h(f_1(0);m=3)-h(f_1(0);m=X)|$ as a function of $f_1(0)$ for several values $X$ (with $\gamma=1$). (b) $\delta(h,\gamma)=|h(f_1(0);\gamma=m)-h(f_1(0);\gamma=1)|$ as a function of $f_1(0)$ for several values of $g(0)$, comparing the two cutoff families $\gamma=1$ and $\gamma=m$.}
\end{figure}

To avoid numerical instabilities close to the divergence in RG time, we reparametrize the flow by using $\lambda$ as the evolution parameter. Concretely, we form ratios such as
\begin{equation}
\frac{d x_0}{d\lambda}=\frac{\dot x_0}{\dot\lambda},\qquad \frac{d g}{d\lambda}=\frac{\dot g}{\dot\lambda},\qquad \frac{d f_1}{d\lambda}=\frac{\dot f_1}{\dot\lambda},\qquad \frac{d w}{d\lambda}=\frac{\dot w}{\dot\lambda},
\end{equation}
and solve the resulting system with initial conditions $x_0(\lambda(0))=x_0(0)$, $g(\lambda(0))=g(0)$, $f_1(\lambda(0))=f_1(0)$, $w(\lambda(0))=w(0)$. For a fixed value of $f_1(0)$ we then scan over $\lambda(0)$ and determine whether the solution reaches $\lambda\to\infty$  at finite $t$ or instead remains regular and flows towards the UV fixed point. The critical value $\lambda(0)=h(f_1(0))$ is identified as the boundary between these behaviors. Repeating this procedure for a range of $f_1(0)$ yields the full curve $\lambda(0)=h(f_1(0))$. The region enclosed by this curve corresponds to the initial conditions for which $\lambda(t)$ becomes asymptotically free in the UV; it is the projection of $M_1$ onto the $(f_1(0),\lambda(0))$ plane at fixed $(x_0(0),g(0),w(0))$. An example of such a projection is shown in Fig.~\ref{plotex}.

In our analysis we find that the value of $x_0(0)$ has no appreciable impact on the critical line $\lambda(0)=h(f_1(0))$. Figure~\ref{plotg} shows $\lambda(0)=h(f_1(0))$ for $m=d/2+1$, $N=4$, $(x_0(0),w(0))=(0.1,1)$ and several values of $g(0)$ chosen near the Gaussian regime $g_\ast=0$. The curves exhibit a single maximum and decay to $\lambda(0)\to 0$ as $f_1(0)\gg 1$. As $g(0)$ increases, the maximum shifts to larger values of $\lambda(0)$. A similar behavior is observed when fixing $g(0)$ and varying $w(0)$: Fig.~\ref{plotw} shows the critical line for different values of $w(0)$ at fixed $g(0)$ and $x_0(0)$. Varying the number of scalar fields $N$ also yields a similar qualitative behavior, as shown in Fig.~\ref{plotN}.

Finally, varying the cutoff-shape parameter $m$ at fixed initial conditions and fixed $N$, we find only a mild dependence of the critical line on $m$. This is illustrated in Fig.~\ref{diffmpl}, which shows the logarithm of the absolute difference
\begin{equation}
\delta(h,X)=\left|h\!\left(f_1(0);m=3\right)-h\!\left(f_1(0);m=X\right)\right|
\end{equation}
for several values $X$ as a function of $f_1(0)$ (with $\gamma=1$). This indicates that the existence and qualitative shape of the critical line are not artifacts of the parameter $m$.

The same analysis can be performed for the second cutoff family $\gamma=m$. For identical initial conditions and the same values of $m$ and $N$, we find that the critical line $\lambda(0)=h(f_1(0))$ is insensitive to the choice of $\gamma$. This is shown in Fig.~\ref{diffgamma}, where we plot
\begin{equation}
\delta(h,\gamma)=\left|h\!\left(f_1(0);\gamma=m\right)-h\!\left(f_1(0);\gamma=1\right)\right|
\end{equation}
as a function of $f_1(0)$ for several values of $g(0)$. We therefore conclude that the critical line $\lambda(0)=h(f_1(0))$ is not an artifact of the cutoff family.

In all plots the critical line starts at a minimum value $f_{1,\min}(0)$ for which $\lambda(0)=0$. This threshold depends only on $N$ and $w(0)$. For $f_1(0)$ slightly above $f_{1,\min}(0)$, the critical line rises approximately linearly in $f_1(0)-f_{1,\min}(0)$, with a slope that depends on $(g(0),w(0),N)$. After reaching a maximum, the curve enters an asymptotic regime in which it decays as
\begin{equation}
\label{critlineasint}
\lambda(0)=\frac{\gamma g_{*}f_{1*}w_*^2}{12(m-1)}\frac{1}{f_1(0)}\qquad \text{for}\quad f_1(0)\gg 1.
\end{equation}
in this regime there is no dependence on the initial conditions of $g(t)$, $x_0(t)$ and $w(t)$. Additionally, there is also no dependence on $m$ since $g_*=\frac{192\pi^2(m-1)}{(16-N)\gamma}$, so the critical line is cutoff independent for $f_1(0)\to\infty$. Also, in this limit, at fixed $N$, all critical lines focus into a single line, this is shown in the inset of fig. \ref{plotg}.

Now we briefly recast the previous results in geometric terms, which also clarifies the origin of the asymptotic relation eq.~(\ref{critlineasint}). The RG flow defines a vector field on the coupling space,
\begin{equation}
\dot{\mathbf{g}}(t)=\boldsymbol{\beta}(\mathbf{g})\,,\qquad 
\mathbf{g}(t)=\big(x_0(t),\lambda(t),g(t),f_1(t),w(t)\big)\,,
\end{equation}
whose integral curves are the RG trajectories. Let $\mathbf{g}_*$ denote the UV fixed point in eq.(\ref{FPbello}) (with the plus sign). The set of initial data at $t=0$ whose trajectories are regular for all $t\ge 0$ and satisfy $\lim_{t\to+\infty}\mathbf{g}(t)=\mathbf{g}_*$ defines the IR UV-complete manifold $M_1$. Equivalently, $M_1$ is the intersection at $t=0$ of the (UV) stable manifold $W^s(\mathbf{g}_*)$ of the fixed point with the IR hyperplane.

Locally, $M_1$ can be described as the zero level-set of a function $S_{\rm reg}(\mathbf{g})$,
\begin{equation}
S_{\rm reg}(\mathbf{g})=0\,,
\end{equation}
and the defining property of such an invariant manifold is that the RG vector field is tangent to it. Denoting by
$\mathbf{n}(\mathbf{g})=\nabla S_{\rm reg}(\mathbf{g})$ the normal to the hypersurface, tangency is expressed by
\begin{equation}
\label{concrit}
0=\mathbf{n}\cdot \boldsymbol{\beta}
=\sum_{i}\,(\partial_{g_i}S_{\rm reg})\,\beta_{g_i}\,,
\qquad g_i\in\{x_0,\lambda,g,f_1,w\}\,.
\end{equation}
Selecting $\lambda$ as a representative coupling, one may equivalently parameterize the manifold as
\begin{equation}
S_{\rm reg}(\mathbf{g})=\lambda-\lambda_{\rm reg}(x_0,g,f_1,w)=0\,,
\end{equation}
so that eq.~(\ref{concrit}) becomes a first-order condition determining $\lambda_{\rm reg}$ along the RG characteristics.

In the large-$f_1(0)$ regime at fixed $N$, the UV-complete trajectories rapidly approach the fixed-point values of the remaining couplings, $(x_0,g,w)\to(x_{0*},g_*,w_*)$, while $f_1(0)$ acts as an external large parameter in the projection to the $(f_1(0),\lambda(0))$ plane. To leading order, the tangency condition eq.~(\ref{concrit}) therefore reduces to the requirement that the $\lambda$-component of the flow vanishes on the asymptotic trajectory \footnote{In eq.~(\ref{condasint}) the coupling $\lambda$ is replaced by $\lambda(0)/2$ and not simply by $\lambda(0)$ due to the Taylor expansion of $u_t(x)$.},
\begin{equation}
\label{condasint}
\beta_\lambda\!\left(x_{0*},\frac{\lambda(0)}{2},g_*,f_1(0),w_*\right)=0\,.
\end{equation}
Solving this equation for $\lambda(0)$ and expanding for $f_1(0)\to\infty$ yields precisely the asymptotic critical line eq.~(\ref{critlineasint}), explaining why, at fixed $N$, the numerical critical lines collapse onto a single curve in that limit (cf.\ the inset of Fig.~\ref{plotg}).

\section{Numerical solutions}\label{sectionnumsol}
\subsection{Numerical runnings}\label{subsecsectionnumsol}
\begin{figure}[t]
         \centering
         \includegraphics[width=0.45\textwidth]{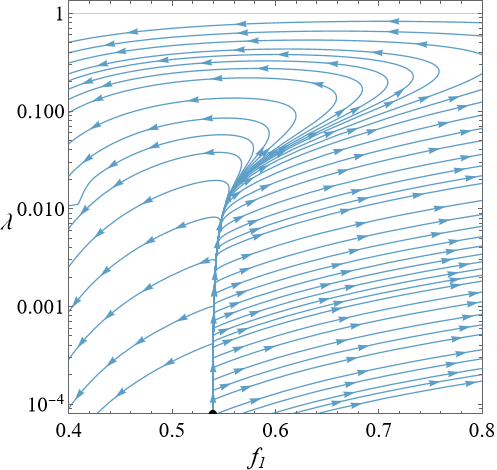}
         \caption{Projection of representative RG trajectories onto the $(f_1,\lambda)$ plane for $t\gtrsim t_{\mathrm{tr}}$ (Planck and UV regime), obtained from initial conditions inside the UV-complete region. The example shown uses $N=4$, $g(0)=10^{-56}$, $w(0)=1$ and $x_0(0)=1$. Arrows indicate the direction of decreasing RG time (from UV to IR). In the Planck regime gravitational fluctuations induce a turnaround: instead of growing monotonically, both $\lambda$ and $f_1$ reach a maximum and then decrease towards the UV fixed point \eqref{FPbello}. For $w(0)=1$ one finds $f_{1\ast}\approx 0.539$, marked by the black dot.}
         \label{plotphase}

\vspace{0.5cm}

    \centering
     \subfigure[]{
         \centering
         \includegraphics[width=0.45\textwidth]{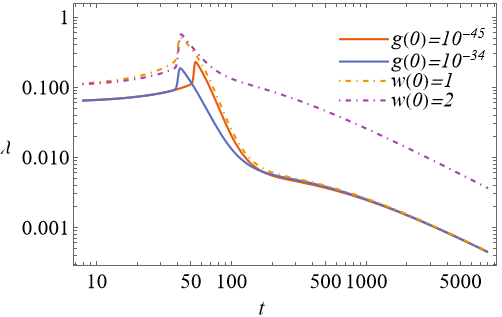}
         \label{plotrunlambda}
     }
    \hspace{0.8em}
     \subfigure[]{
         \centering
         \includegraphics[width=0.45\textwidth]{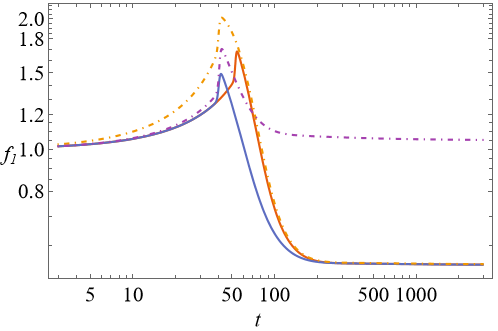}
         \label{plotrunf1}
     }
     \caption{Representative UV-complete runnings for $N=4$. (a) Running of $\lambda(t)$. (b) Running of $f_1(t)$. Solid lines: different values of $g(0)$ at fixed $w(0)$ and $x_0(0)$. Dot-dashed lines: different values of $w(0)$ at fixed $g(0)$ and $x_0(0)$.}
\end{figure}

\noindent The flow equations in Appendix~\ref{appbeta} can be solved only numerically. Eliminating $t$ from the solutions, one can construct RG trajectories in the five-dimensional coupling space $(x_0,\lambda,g,f_1,w)$ or equivalently the hypersurface $\mathrm{S}_{reg}$. Since these trajectories cannot be visualized in the full space, we show representative projections of $\mathrm{S}_{reg}$. In particular, the projection onto the $(f_1,\lambda)$ plane for scales above the Planck regime is shown in Fig.~\ref{plotphase}. For initial conditions inside the UV-complete region (the projection of $M_1$ in Fig.~\ref{plotex}), all trajectories approach the interacting UV fixed point.

Before presenting the numerical runnings, it is useful to describe how to fix the boundary conditions at $t=0$, especially for the Newtonian constant and for the minimum of the potential. In the IR region at $t=0$ we require that
\begin{equation}
G(t=0)=\frac{1}{m_p^2}
\end{equation}
where $m_p$ is the Plank mass: $m_p\simeq 10^{19}\,\mathrm{GeV}$. In terms of adimensional Newtonian constant, this translates to
\begin{equation}
g(t=0)=\frac{\Lambda_0^2}{m_p^2}.
\end{equation}
In the IR (namely, for the solutions where $t\le0$), gravitational fluctuations should decouple from the dynamics and this restricts $g(0)$ to lie in the Gaussian regime so that the relation 
\begin{equation}
\label{natcondx0}
\Lambda_0\ll m_p
\end{equation}
has to be satisfied to have a well-defined decoupling regime in which gravitational fluctuations do not dominate the infrared dynamics. In this regime, infrared observables are expected not to exhibit large hierarchies induced by the ultraviolet scale. In particular, this suggests that the IR value of $x_0$ is of order one. We therefore choose the representative value
\begin{equation}
x_0(t=0)=Z_L(\Lambda_0)\frac{\rho_0}{\Lambda_0^2}=1,
\end{equation}
that is, the vacuum expectation value should be of the same order as the measurement scale $\Lambda_0$. The condition $x_0(0)=1$ has to be understood as a representative convenient choice within the natural class of IR configurations that do not exhibit fine-tuning and not as an UV-motivated constraint. In particular, the numerical analysis on the UV complete trajectories shows that the runnings are not affected by the initial condition on $x_0$ so that a unique $x_0(0)$ can be used to describe the runnings.

The RG-time evolution of representative UV-complete trajectories for $N=4$ is shown in Figs.~\ref{plotrunlambda} and \ref{plotrunf1}. Solid lines correspond to solutions at fixed $w(0)$ and $x_0(0)$ for two different values of $g(0)$, namely $g(0)=10^{-45}$ and $g(0)=10^{-34}$. Dot-dashed lines correspond to solutions at fixed $g(0)$ and $x_0(0)$ for two different values of $w(0)$, namely $w(0)=1$ and $w(0)=2$. The running $\lambda(t)$ follows the matter-dominated (Gaussian)
behavior up to a transition value
\begin{equation}
t_{\mathrm{tr}}\sim \ln\!\left(\frac{m_p}{\Lambda_0}\right),
\end{equation}
at which gravitational fluctuations become important and $g(t)$ crosses over from the Gaussian regime to the non-Gaussian fixed-point regime. Around this crossover, $\lambda(t)$ reaches a maximum and subsequently decreases towards $\lambda_\ast=0$, with an asymptotic behavior compatible with $\lambda(t)\sim 1/t$ at large $t$. The running of $f_1(t)$ shows a similar crossover and approaches its fixed-point value $f_{1\ast}$ in the UV. 

The transition scale is where the critical condition eq.(\ref{critval}) starts to be satisfied. Indeed, the maximum of $\lambda(t)$ corresponds to a zero of the beta function $\beta_\lambda$. This is exactly where the condition $\lambda_I=A_I/B_I$ of Sec. \ref{sectionbeta} holds and then for larger $t$ the inequality $|A(t)|>B(t)\lambda(t)$ is satisfied, allowing the decreasing of $\lambda(t)$. In particular, in the limit $t=+\infty$ eq.(\ref{critval}) gives the constraint $1/g_{*}=f_{1*}x_{0*}$, which eq.(\ref{FPbello}) satisfies, and the running of $f$ and $u$ reduces to $f(t\to\infty)=f_{1*}x$ and $u(t\to\infty)=u_{0*}$ reproducing the exact scaling solution. 

\begin{figure}[p]
     \centering
     \subfigure[]{
         \centering
         \includegraphics[width=0.45\textwidth]{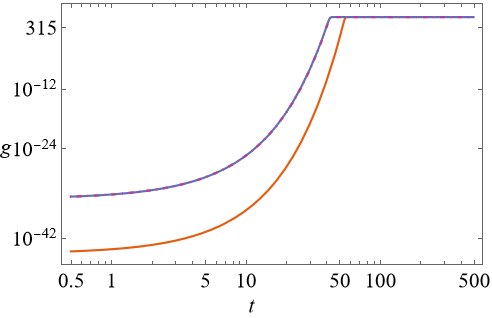}
         \label{plotrung}
     }
    \hspace{0.8em}
     \subfigure[]{
         \centering
         \includegraphics[width=0.45\textwidth]{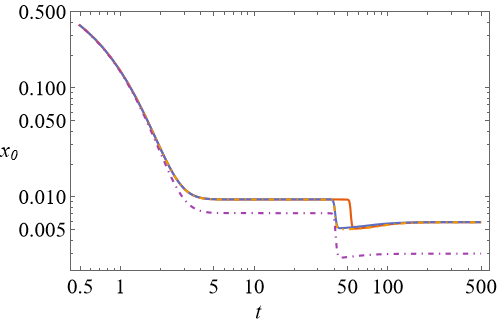}
         \label{plotrunx0}
     }
     \caption{Representative UV-complete runnings for $N=4$. (a) Running of $g(t)$, showing the crossover from the Gaussian regime to the non-Gaussian scaling regime around $t_{\mathrm{tr}}$. (b) Running of $x_0(t)$, which approaches a quasi-constant value in the fixed-point regime of Eq.~(\ref{FPbello}).}

\vspace{0.5cm}

     \centering
     \subfigure[]{
         \centering
         \includegraphics[width=0.45\textwidth]{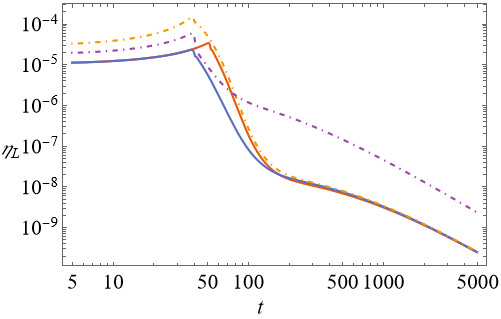}
         \label{plotrunetaL}
     }
    \hspace{0.8em}
     \subfigure[]{
         \centering
         \includegraphics[width=0.45\textwidth]{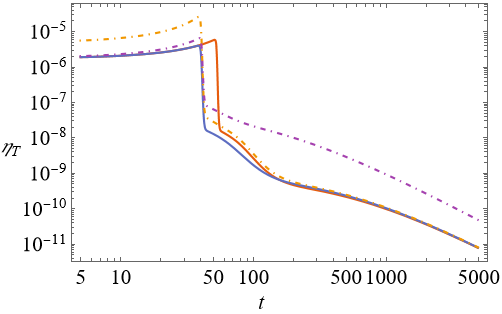}
         \label{plotrunetaT}
     }
     \caption{Representative UV-complete runnings for $N=4$. (a) Running of $\eta_L(t)$. (b) Running of $\eta_T(t)$. Both anomalous dimensions peak around the crossover and then decay slowly towards $\eta_{L\ast}=0$ and $\eta_{T\ast}=0$}

\vspace{0.5cm}

     \centering
     \subfigure[]{
         \centering
         \includegraphics[width=0.45\textwidth]{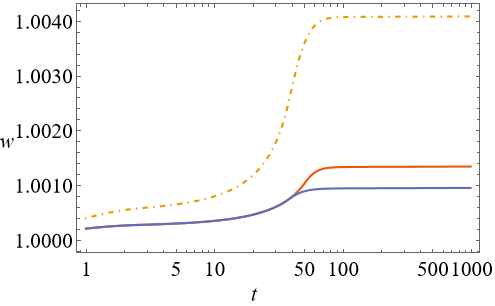}
         \label{plotrunw1}
     }
    \hspace{0.8em}
     \subfigure[]{
         \centering
         \includegraphics[width=0.45\textwidth]{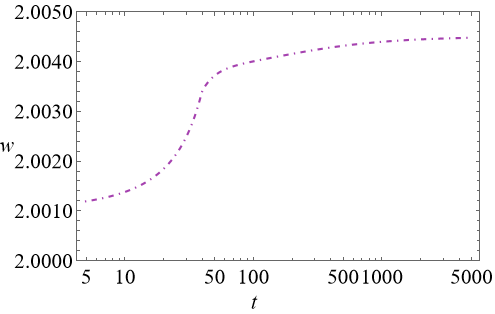}
         \label{plotrunw2}
     }
     \caption{Representative UV-complete runnings for $N=4$. (a) Running of $w(t)$ for a solution with $w(0)=1$. (b) Running of $w(t)$ for a solution with $w(0)=2$. In both cases $w(t)$ stays close to $w(0)$ in the Gaussian regime and approaches its UV value after the crossover. The IR initial conditions fix the free parameter $w_\ast$ of the UV fixed point eq.(\ref{FPbello}).}
\end{figure}

In all cases, the running of $g(t)$ shown in Fig.~\ref{plotrung} is qualitatively similar to the pure-gravity behavior: a Gaussian regime at low scales crosses over to the non-Gaussian scaling regime around $t_{\mathrm{tr}}$. The running of $x_0(t)$, shown in Fig.~\ref{plotrunx0}, exhibits two quasi-constant regimes; the second is approached in the non-Gaussian fixed-point regime of Eq.~(\ref{FPbello}). In particular, the system remains in the symmetry-broken phase along the displayed trajectories.

Figures~\ref{plotrunetaL} and \ref{plotrunetaT} show the anomalous dimensions $\eta_L(t)$ and $\eta_T(t)$. Both remain small but non-vanishing. As for $\lambda(t)$, they develop a peak around the crossover to the non-Gaussian regime and then decay slowly towards their fixed-point values $\eta_{L\ast}=0$ and $\eta_{T\ast}=0$. Due to the smallness of $\eta_{L,T}$, the ratio $w(t)$ in Figs.~\ref{plotrunw1} and \ref{plotrunw2} is nearly constant over the full RG interval. Deviations from the IR value $w(0)$ start around $t_{\mathrm{tr}}$, after which $w(t)$ approaches its UV value. The initial conditions thus fix the free parameter $w_\ast$ of the UV fixed point eq.~(\ref{FPbello}), which in turn determines the corresponding fixed-point values $x_{0\ast}(w_\ast)$ and $f_{1\ast}(w_\ast)$.

Changing $g(0)$ at fixed $w(0)$ and $x_0(0)$ primarily affects the length of the matter-dominated regime: the smaller $g(0)$ is, the longer the trajectory stays close to the Gaussian regime before crossing over. The location of the maximum in $\lambda$ and $f_1$ shifts accordingly, while the UV scaling behavior is essentially unchanged. By contrast, changing $w(0)$ at fixed $g(0)$ and $x_0(0)$ mainly affects the UV scaling regime and its approach to the fixed point. While $g(t)$ depends only weakly on $w(0)$, the maxima of $\lambda(t)$ and $f_1(t)$, as well as the asymptotic coefficients in the large-$t$ expansions (e.g.\ $\lambda(t)\sim 1/t+O(1/t^2)$), show a pronounced dependence on $w(0)$. This is reflected in the different numerical values of
$x_{0\ast}(w_\ast)$ and $f_{1\ast}(w_\ast)$ approached in the UV.

\subsection{Limits on  the mass of the scalar}
The independence of the critical line $\lambda(0)=h\left(f_1(0)\right)$ from $m$ and $\gamma$ allows us to extract predictions that are insensitive to the cutoff choice. In particular, it can be used to constrain the scalar mass in the broken phase, cf.\ Eq.~(\ref{massL}), evaluated at the reference scale $t=0$. 

\begin{figure}[t]
     \centering
     \subfigure[]{
         \centering
         \includegraphics[width=0.45\textwidth]{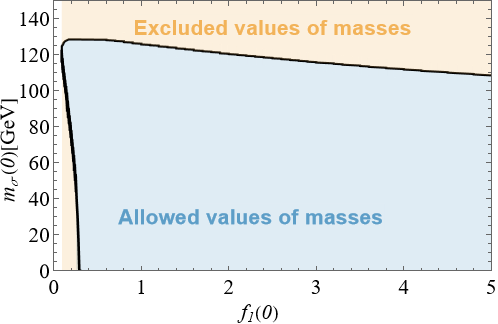}
         \label{plotmsex}
     }
    \hspace{0.8em}
     \subfigure[]{
         \centering
         \includegraphics[width=0.45\textwidth]{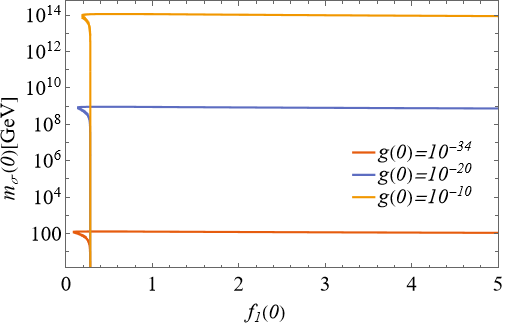}
         \label{plotmsg}
     }
     \caption{Constraints on the scalar mass from the UV-complete region in the $(f_1(0),m_\sigma(0))$ plane. (a) Example at $g(0)=10^{-34}$, $x_0(0)=1$, $w(0)=1$ and $N=4$: the light-blue region corresponds to the projection of the IR critical manifold $M_1$ (UV-complete trajectories), while the beige region corresponds to singular trajectories. The black curve is the critical mass line separating the two behaviors. (b) Same as (a), shown for several values of $g(0)$ (logarithmic scale).}

\vspace{0.5cm}

     \centering
     \subfigure[]{
         \centering
         \includegraphics[width=0.45\textwidth]{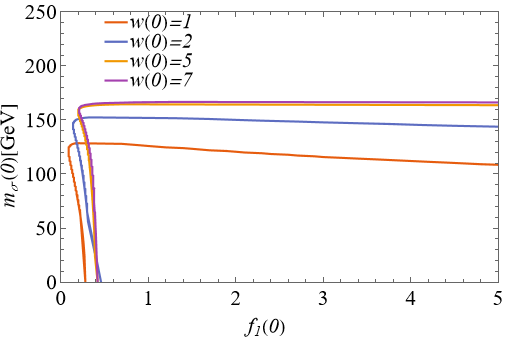}
         \label{plotmsw}
     }
    \hspace{0.8em}
     \subfigure[]{
         \centering
         \includegraphics[width=0.45\textwidth]{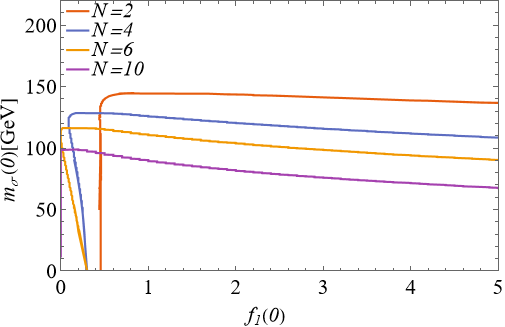}
         \label{plotmsN}
     }
     \caption{Dependence of the critical mass line on IR parameters. (a) Variation with $w(0)$ at fixed $N=4$ and fixed $(g(0),x_0(0))$. (b) Variation with $N$ at fixed $(g(0),x_0(0),w(0))$.}
\end{figure}

Figure~\ref{plotmsex} illustrates how the projection of the IR critical manifold constrains the scalar mass at fixed $g(0)$, $w(0)$ and $N$. In this representation, the critical line in the $(f_1(0),\lambda(0))$ plane becomes a \emph{critical mass line} in the $(f_1(0),m_\sigma(0))$ plane: as in the previous discussion, the region below the curve corresponds to UV-complete trajectories, while points above it lead to Landau-pole-like behavior. The critical mass line exhibits a maximum and then decreases at large $f_1(0)$. In the asymptotic limit $f_1(0)\gg 1$, using eq.(\ref{critlineasint}) we get 
\begin{equation}
m_{\sigma,\mathrm{crit}}(f_1(0)\to\infty)=\frac{m_p(0)}{3}\sqrt{\frac{\gamma g_\ast f_{1\ast}}{2\left(m-1\right)\left(N-1\right)}}w_\ast\left(\frac{1}{f_1\left(0\right)}\right)^{3/2}+O\left(\frac{1}{f_1\left(0\right)}\right)^{5/2}
\end{equation}
replacing the fixed point value $g_*=\frac{192\pi^2(m-1)}{(16-N)\gamma}$ shows that this expression does not depend on $m$, as for the asymptotic of the critical line. Furthermore, in this asymptotic limit, all critical mass lines focus on a single critical mass line.

Figure~\ref{plotmsg} shows the critical mass line as a function of $f_1(0)$ for $N=4$ and several values of $g(0)$. The allowed mass range depends strongly on $g(0)$, changing by many orders of magnitude as $g(0)$ is varied. By contrast, at fixed $g(0)$, the dependence on $w(0)$ (Fig.~\ref{plotmsw}) and on $N$ (Fig.~\ref{plotmsN}) is milder, and the resulting mass range typically remains within the same order of magnitude.

The maximum allowed scalar mass is determined by the peak of the critical line, i.e.\ by the pair $\bigl(\bar f_1(0),\lambda_{\max}(0)\bigr)$ with $\lambda_{\max}(0)=h\!\left(\bar f_1(0)\right)$. In $d=4$, at $t=0$, one finds
\begin{equation}
m_{\sigma,\max}(t=0)=2\Lambda_0\sqrt{\frac{\lambda_{\max}(0)\,x_0(0)}{1+6\,\bar f_1^{\,2}(0)\,x_0(0)\,g(0)}}\, .
\end{equation}
This expression makes the strong sensitivity to $g(0)$ explicit. Expanding around $g(0)=0$ yields $m_{\sigma,\max}(t=0)=2\Lambda_0\sqrt{x_0(0)\lambda_{\max}(0)}+O\!\left(g(0)\right)$. Since $\Lambda_0^2=m_p^2 g(0)$, the overall scale is set by the competition between the largeness of $m_p$ and the smallness of $g(0)$. If $g(0)$ is not small enough to compensate the Planck scale, the maximum scalar mass remains parametrically below $m_p$; conversely, for extremely small $g(0)$ the maximum mass becomes very small. For example, for $g(0)=10^{-56}$ (corresponding to a measurement performed at $\Lambda_0\sim 1\,\mathrm{eV}$) we find $m_{\sigma,\max}(t=0)\sim 10^{-10}\,\mathrm{GeV}$ but with $g(0)=10^{-10}$  (corresponding to $\Lambda_0\sim 10^{9}\,\mathrm{GeV}$) we find $m_{\sigma,\max}(t=0)\sim 10^{14}\,\mathrm{GeV}$.

As a further illustrative example, consider $N=4$ and a reference scale near the top-quark mass, $\Lambda_0=m_t\simeq 173\,\mathrm{GeV}$, for which $g(0)\sim 10^{-34}$. In this case we find $\lambda_{\max}(0)\simeq 0.138$, which along with the condition $x_0(0)=1$ gives $v\simeq Z_L(\Lambda_0)244\mathrm{GeV}$ and then corresponds to $m_{\sigma,\max}\simeq 128.5\,\mathrm{GeV}$. 

Following the UV complete trajectories from the UV scaling regime ($t\gg t_{\mathrm{tr}}$) to the IR ($t\ll t_{\mathrm{tr}}$) can produce scalar masses both much larger and much smaller than the Planck mass, depending on how the parameters in the UV critical manifold evolve in the IR. This happens because although for $t\ll t_{\mathrm{tr}}$, gravitational fluctuations decouple, so that infrared observables are not driven by the ultraviolet scale and do not develop UV-induced hierarchies, this does not preclude hierarchies that originate from the choice of UV data along relevant directions. In particular, the constraint $m_\sigma(t=0) \ll m_p$ remains sensitive to the UV initial conditions. Accordingly, achieving scalar masses in the IR much smaller than the Planck scale while simultaneously matching $G(t=0)=1/m_p^2$ generally requires a degree of fine-tuning in the UV initial conditions. In contrast, the choice $x_0(0)=1$ does not follow from the dynamics but represents a convenient IR reference value within the class $x_0(0)=\mathcal{O}(1)$.

\subsection{Running of the scalar mass}
\begin{figure}[t]
     \centering
     \subfigure[]{
         \centering
         \includegraphics[width=0.45\textwidth]{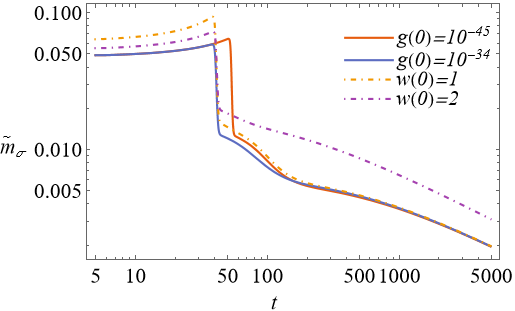}
         \label{plotmassadim}
     }
    \hspace{0.8em}
     \subfigure[]{
         \centering
         \includegraphics[width=0.45\textwidth]{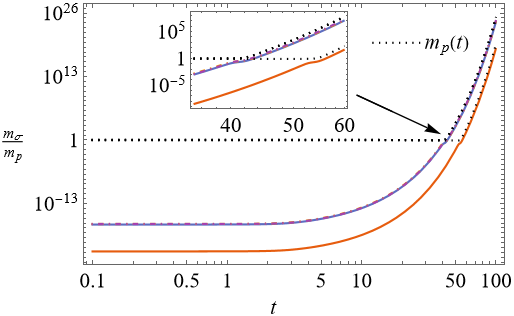}
         \label{plotmassdim}
     }
     \caption{Running of the longitudinal scalar mass for the four solutions discussed in the subsection \ref{subsecsectionnumsol}. In (a) the running of dimensionless mass. In (b) the running of dimensionfull mass in unit of Plank mass. The zoom around the transition scale highlights the change of scaling behavior from $m_\sigma(t)\sim m_\sigma(0)e^{t}$ to $\sim m_\sigma(0)t^{-1/2}e^{t}$. The black dotted lines show the running of Plank mass. Above the transition scale there is no distinction between the running of $m_p$ and $m_\sigma$.}
\end{figure}

We now discuss the running of the longitudinal scalar mass. Figure~\ref{plotmassadim} shows the RG evolution of the
dimensionless mass $\tilde m_\sigma$ for the four numerical solutions introduced in Sec.~\ref{subsecsectionnumsol}. For
$t\ll t_{\rm tr}$ the running is almost frozen, $\tilde m_\sigma(t)\simeq \tilde m_\sigma(0)$. Around the transition scale
$t_{\rm tr}$, where gravitational fluctuations start to dominate, $\tilde m_\sigma(t)$ develops a maximum and then crosses a
short transient regime before approaching the UV fixed point value $\tilde m_{\sigma*}=0$.

In the asymptotic UV regime we find a slow decay,
\begin{equation}
\tilde m_\sigma(t)\sim t^{-1/2}\qquad (t\to\infty).
\end{equation}
This behavior is expected: near the UV fixed point all couplings approach constant values, while the quartic coupling is
marginally running as $\lambda(t)\sim 1/t$. Since at leading order $\tilde m_\sigma^2\propto x_0\,\lambda$ (cf.\
Eq.~\eqref{massL}), one obtains $\tilde m_\sigma(t)\propto \sqrt{\lambda(t)}\sim t^{-1/2}$.

The dependence on the IR initial conditions mirrors the behavior discussed for $\lambda(t)$ in Sec.~\ref{subsecsectionnumsol}:
changing $g(0)$ at fixed $w(0)$ mainly shifts the extent of the matter-dominated regime below $t_{\rm tr}$, whereas changing
$w(0)$ at fixed $g(0)$ primarily affects the gravity-dominated regime above $t_{\rm tr}$.

Figure~\ref{plotmassdim} shows the corresponding dimensionful mass, plotted in units of the reference Planck mass at $t=0$.
Below $t_{\rm tr}$ the nearly constant $\tilde m_\sigma$ implies the canonical scaling $m_\sigma(t)\propto \Lambda\propto e^t$.
Above $t_{\rm tr}$ the UV decay of $\tilde m_\sigma$ induces an additional suppression, so that
\begin{equation}
m_\sigma(t)=\Lambda\,\tilde m_\sigma(t)\;\sim\; e^{t}\,t^{-1/2}\,,
\qquad (t\gtrsim t_{\rm tr}) .
\end{equation}
In the same figure we also show the running Planck mass $m_p(t)$ (dotted curves). In the matter-dominated regime $m_p(t)$ is
approximately constant, while above $t_{\rm tr}$ it follows the same overall scaling pattern as $m_\sigma(t)$. This can be
understood from the critical relation eq.~(\ref{critval}), which becomes approximately saturated above the transition and implies
$x_0\sim 1/(f_1 g)$. Inserting this into Eq.~(\ref{massL}) yields the parametric estimate
\begin{equation}
m_\sigma(t)\;\sim\;2\,m_p(t)\,\sqrt{\frac{\lambda(t)}{f_1(t)\,\bigl(1+6f_1(t)\bigr)}}\,,
\end{equation}
showing that $m_\sigma$ is of order $m_p$ up to a factor $\propto \sqrt{\lambda(t)}$, and hence $m_\sigma/m_p\to 0$ as
$\lambda\to 0$ in the UV.

\subsection{The numerical runnings in the deep IR regime}
The numerical integration can also be extended to $t \leq 0$, corresponding to the deep infrared (IR) regime. In this region, gravitational fluctuations effectively decouple, and the running of the couplings becomes insensitive to $g(t)$. The dynamics is then entirely driven by matter fluctuations. In particular, the flows of $x_0(t)$, $\lambda(t)$ and $w(t)$ decouple from the flow of $f_1(t)$, so that $f_1(t)$ is fully determined by the running of $x_0(t)$, $\lambda(t)$ and $w(t)$. For fixed initial conditions on $\lambda(0)$, the residual dependence of the flow is controlled by the choice of $w(0)$.

In order to display the infrared behavior in a log-log representation, we redefine the RG time as $\tau = -t$, such that the deep IR limit $t \to -\infty$ corresponds to $\tau \to +\infty$. This reparametrization is introduced purely for visualization purposes.

\begin{figure}[p]
     \centering
     \subfigure[]{
         \centering
         \includegraphics[width=0.45\textwidth]{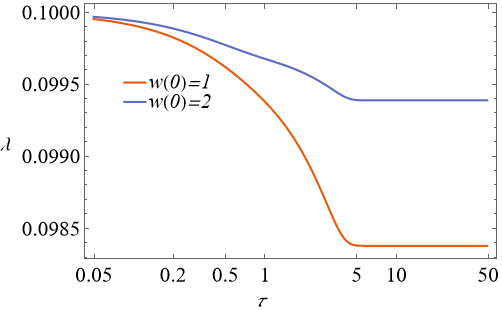}
         \label{plotrunlambdatneg}
     }
    \hspace{0.8em}
     \subfigure[]{
         \centering
         \includegraphics[width=0.45\textwidth]{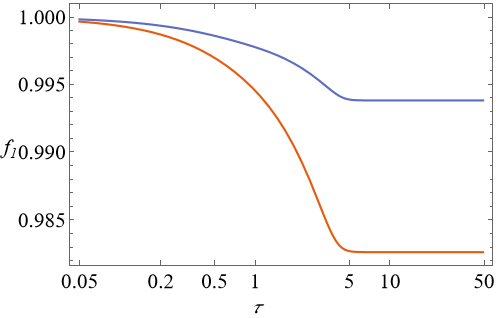}
         \label{plotrunf1tneg}
     }
     \caption{Representative runnings in the negative-$t$ region ($\tau = -t$) for $\lambda(\tau)$ (a) and $f_1(\tau)$ (b), corresponding to the solutions shown in Figs.~\ref{plotrunlambda} and \ref{plotrunf1} with $w(0)=1$ and $w(0)=2$.}

\vspace{0.5cm}

     \centering
     \subfigure[]{
         \centering
         \includegraphics[width=0.45\textwidth]{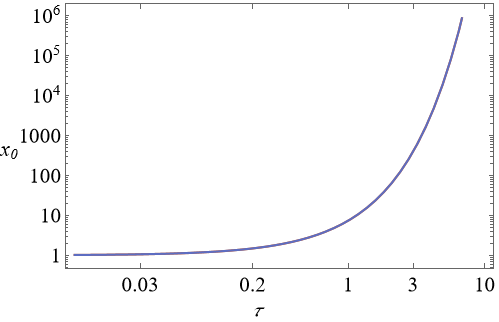}
         \label{plotrunx0tneg}
     }
    \hspace{0.8em}
     \subfigure[]{
         \centering
         \includegraphics[width=0.45\textwidth]{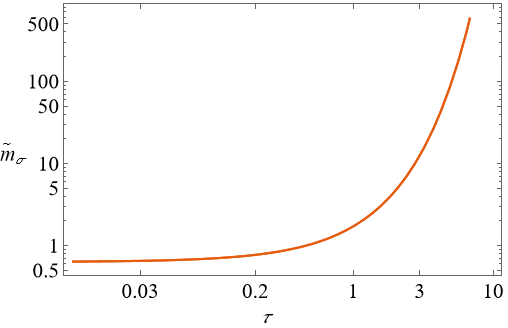}
         \label{plotrunmstneg}
     }
     \caption{Representative runnings in the negative $t$ region ($\tau=-t$) for $x_0(\tau)$ (a) and $\tilde m_s(\tau)$ (b) corresponding to the solutions in figs. \ref{plotrunx0} and \ref{plotmassadim} with $w(0)=1$ and $w(0)=2$. In contrast to the other runnings, the solutions show no visible dependence on the initial value $w(0)$.}

\vspace{0.5cm}

     \centering
     \subfigure[]{
         \centering
         \includegraphics[width=0.45\textwidth]{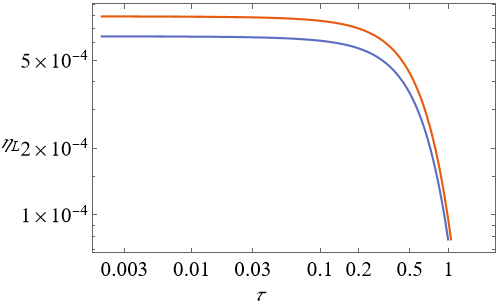}
         \label{plotrunetaLtneg}
     }
    \hspace{0.8em}
     \subfigure[]{
         \centering
         \includegraphics[width=0.45\textwidth]{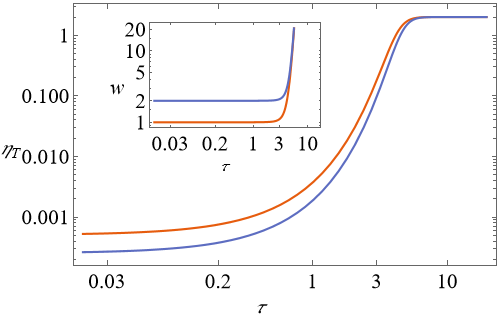}
         \label{plotrunetaTtneg}
     }
     \caption{Representative runnings in the negative $t$ region ($\tau=-t$) for $\eta_L(\tau)$ (a) and $\eta_T(\tau)$ (b) corresponding to the solutions in figs. \ref{plotrunetaL} and \ref{plotrunetaT} with $w(0)=1$ and $w(0)=2$. In the inset of (b) the corresponding $w(\tau)$ is shown.}
\end{figure}

Figs.~\ref{plotrunlambdatneg} and \ref{plotrunf1tneg} show the numerical running of $\lambda(t)$ and $f_1(t)$ for $t \leq 0$, corresponding to the solutions with $w(0)=1$ and $w(0)=2$ shown in Figs.~\ref{plotrunlambda} and \ref{plotrunf1}. Both couplings decrease and approach asymptotic values in the deep IR ($t \ll 0$), with limiting values that depend on $w(0)$.

In the deep IR, the running minimum enters the Gaussian regime and exhibits canonical scaling, $x_0(t \ll 0) \sim e^{-2t}$ (Fig.~\ref{plotrunx0tneg}). Similarly, the dimensionless scalar mass scales as $\tilde m_\sigma(t \ll 0) \sim e^{t}$ (Fig.~\ref{plotrunmstneg}). This corresponds to the system flowing into the spontaneously broken phase, where the dimensionful vacuum expectation value approaches a constant. In contrast to the running of $\lambda(\tau)$ and $f_1(\tau)$, the solutions show no visible dependence on the initial value $w(0)$: the corresponding trajectories are numerically indistinguishable within the resolution of the plots.

The behavior of the anomalous dimensions in the deep IR clarifies the effective dynamics of the model. The longitudinal anomalous dimension $\eta_L$ (Fig.~\ref{plotrunetaLtneg}) vanishes for $t \ll 0$, while the transverse anomalous dimension approaches $\eta_T \to 2$ (Fig.~\ref{plotrunetaTtneg}). Accordingly, the wavefunction renormalizations scale as $Z_L \sim \text{const}$ and $Z_T \sim e^{-2t}$. In this regime, the longitudinal (radial) mode decouples from the flow, and the dynamics is dominated by the $N-1$ transverse (Goldstone) modes. As a result, the effective IR theory is governed by Goldstone fluctuations and is equivalent to a non-linear sigma model at scales $\Lambda \ll m_\sigma$. This behavior is well established within the functional renormalization group framework for $O(N)$ models \cite{Berges:2000ew,Dupuis:2020fhh,Delamotte:2007pf,WETTERICH1991529}.

The dominance of Goldstone modes is also reflected in the behavior of $w(t)$ (inset of Fig.~\ref{plotrunetaTtneg}), which in the deep IR scales as
\begin{equation}
w(t \ll 0) \sim e^{-2t},
\end{equation}
indicating that $Z_T(t \to -\infty) \gg Z_L(t \to -\infty)$.

The behavior $\eta_T \simeq 2$ should not be interpreted as a physical anomalous dimension, but rather as a consequence of the strong running of the wavefunction renormalization within the present truncation. In particular, the scaling $Z_T(\Lambda) \sim \Lambda^{-2}$ implies that the combination $Z_T \Lambda^2$ approaches a constant, consistently with the emergence of Goldstone dynamics in the infrared. In more refined treatments, the anomalous dimension of Goldstone modes is known to remain small, so $\eta_T \simeq 2$ should be regarded as a truncation artifact \cite{Dupuis:2020fhh,Delamotte:2007pf}.

Although the dimensionless couplings $\lambda$ and $f_1$ approach constant values in the deep IR, this does not correspond to a finite interaction strength for canonically normalized Goldstone fields. Due to the strong running of the transverse wavefunction renormalization $Z_T$, the corresponding physical interactions are suppressed in the infrared.

\section{Conclusions}\label{sectionconclusion}
\noindent In this work we studied an $O(N)$ scalar theory non-minimally coupled to Einstein gravity in the symmetry-broken phase. Using a polynomial truncation around the running minimum of the scalar potential, we employed the proper-time flow equation to compute the RG running of the minimum of the potential, the non-minimal coupling, the quartic coupling, and the cosmological-constant term within the truncation.

Working with the exponential parametrization and the physical gauge, we identified an interacting UV-attractive fixed point. This fixed point opens the possibility of UV-complete trajectories for the quartic scalar coupling without invoking fermionic degrees of freedom. The fixed point is associated with a scaling solution of the underlying flow equation of the form $u=u_\ast$ and $f=f_{1\ast}x$. Within the truncation scheme for the derivative expansion adopted in this work, this scaling solution is realized exactly at the fixed point and is not affected by the order of the polynomial expansion around the running minimum, as well as on the cutoff parameter $m$. As shown in Appendix \ref{apppert}, these properties follow from an analysis of the linearized flow around the scaling solution, which does not rely on a finite polynomial truncation. In particular, this perturbative analysis leads to a consistent determination of the UV critical manifold. This independence from the polynomial truncation and $m$ provides non-trivial evidence for the robustness of the corresponding UV scaling regime, although a full assessment of truncation effects would require more refined approximations.

The fixed-point value of $f_{1\ast}$ lies close to the conformal value, with $f_{1\ast}$ determined by Eq.~(\ref{FPbello}). Closely related fixed-point structures in scalar--tensor systems have also been reported within the Wetterich formalism, see e.g.~\cite{Labus:2015ska}.

The UV fixed point spans a finite-dimensional UV critical surface. Its stability spectrum contains the canonical exponents $\theta=0$, $\theta=2$ and $\theta=4$, as well as additional non-trivial exponents $\theta_{5,6}$ that depend on $N$ and on the free fixed-point parameter $w_\ast$ (see Figs.~\ref{plottheta1} and \ref{plottheta2}). The associated eigendirections are, in general, non-trivial superpositions in the space of couplings. Apart from the direction associated with $u_0$ (with $\theta=4$), the most relevant deformation has a dominant projection along the Newton coupling $g$. This indicates that the approach to the UV fixed point is largely controlled by gravitational fluctuations. In particular, the fixed point breaks the classical marginality of $f_1$ and renders it relevant, while $\lambda$ is marginal at linear order and acquires logarithmic corrections beyond linear order, consistent with an asymptotic behavior of the form $\lambda(t)\sim 1/t+O(1/t^2)$.

Evolving trajectories on the UV critical surface towards the infrared defines a corresponding basin of attraction in the space of IR couplings. This basin contains precisely those IR initial conditions that lead to UV-complete trajectories attracted to the UV fixed point. In practice, this separates IR initial data into two classes: one class yields regular flows for all scales and approaches the UV fixed point, while the complementary class develops a singularity at finite $t$. The separating boundary can be characterized numerically and, at fixed $(g(0),w(0),x_0(0),N,m)$, appears as a critical line $\lambda(0)=h(f_1(0))$ in the $(f_1(0),\lambda(0))$ plane (see Fig.~\ref{plotex}). We find that this critical line is robust under changes of the cutoff family (switching between $\gamma=1$ and $\gamma=m$) and depends only mildly on the cutoff-shape parameter $m$. Its dependence on $x_0(0)$ is also weak, while the strongest dependence is on $g(0)$, $w(0)$ and $N$, in line with the fact that the UV scaling regime is strongly influenced by the $g$-direction.

The critical line provides a practical UV-completion criterion: for given $(g(0),w(0))$, UV-complete trajectories correspond to IR values $(f_1(0),\lambda(0))$ lying in the region bounded by $\lambda(0)=h(f_1(0))$. Adopting the naturalness-motivated choice $x_0(0)=1$ allows us to translate this constraint into bounds on the physical mass of the longitudinal mode $m_\sigma(t=0)$ via Eq.~(\ref{massL}). Interestingly enough, the peak of the critical line determines an upper bound $m_{\sigma,\max}(t=0)$ for fixed $g(0)$ and $N$. For illustrative choices of $\Lambda_0$ in the vicinity of the electroweak scale, the resulting upper bounds can lie in the $\mathcal{O}(10^2\,\mathrm{GeV})$ range.

In the deep infrared ($t\ll0$), the flow exhibits the characteristic features of the spontaneously broken phase: the radial mode becomes effectively decoupled, while the transverse Goldstone modes dominate the dynamics. In this regime, the effective theory is well described by a non-linear sigma model, with the radial fluctuations frozen out and the dynamics constrained to the vacuum manifold.

An important next step is to extend the present framework to more realistic matter sectors. Including gauge fields and fermions introduces additional fluctuation channels that feed into the running of the scalar potential and of the non-minimal coupling. In particular, gravitational fluctuations are known to favor antiscreening contributions in Abelian gauge sectors, suggesting that the mechanism identified here may persist in the presence of gauge interactions and could potentially be relevant for curing Landau-pole behavior in Abelian theories. At the same time, Yukawa interactions provide a competing contribution to the scalar self-coupling and can shift the UV behavior of $\lambda$. It will therefore be important to determine whether the UV-attractive fixed point characterized here by $f_{1\ast}>0$ and $\lambda_\ast=0$ survives once gauge and Yukawa couplings are included, and to map the resulting UV-complete region in the enlarged theory space.

The present analysis is formulated within a Wilsonian renormalization group framework based on a proper-time flow equation. In this setting, the proper-time construction can be interpreted as defining a consistent Wilsonian flow associated with a specific coarse-graining prescription. At the same time, closely related structures are expected to arise within the Wetterich formulation of the functional renormalization group for the effective average action. In comparable truncations and for consistent choices of gauge and field parametrization, both approaches are known to exhibit similar fixed-point structures. For this reason, it will be important to repeat and cross-check the present analysis within an exact EAA/Wetterich framework, in order to further assess the robustness of the results.

\section*{Acknowledgements}
We thank Andrea Spina  Gabriele Giacometti for useful discussions, and Dario Zappalà and Giampaolo Vacca for comments on the manuscript. A.B. is grateful to Holger Gies for enlightening discussions on and to the Physics Department of the University of Jena for hospitality.
\newpage

\begin{appendices}
\section{Derivation of the flow equations for \texorpdfstring{$Z_L$}{ZL} and \texorpdfstring{$Z_T$}{ZT}}
\label{appZLZT}

\noindent In this appendix we derive the proper-time flow equations for the running wave-function renormalizations $Z_L(\Lambda)$ and $Z_T(\Lambda)$ entering Eq.~(\ref{theorysemk}). We build on the background-field and heat-kernel setup of Ref.~\cite{Bonanno:2025qsc}.

\subsection{Projection on the kinetic operators}
To extract $\Lambda\partial_\Lambda Z_T(\Lambda)$ and $\Lambda\partial_\Lambda Z_L(\Lambda)$ we project the flow on the operators $\phi_a(-\Box)\phi_a$ ($a=1,\dots,N-1$) and $\phi_N(-\Box)\phi_N$, respectively. Since we are projecting on the scalar kinetic terms, only the scalar block of the Hessian is required. The projection reads
\begin{equation}
\label{projZ}
\int d^dx\sqrt g\left[\frac{1}{2}\left(\Lambda\partial_\Lambda Z_T\right)\sum_{a=1}^{N-1}\phi_a(-\Box)\phi_a+\frac{1}{2}\left(\Lambda\partial_\Lambda Z_L\right)\phi_N(-\Box)\phi_N\right]
=\frac{1}{2}\int_{0}^{\infty}\frac{ds}{s}\,r\!\left(s,\Lambda^2 \mathbf{Z}\right)\,
\mathrm{Tr}\!\left[e^{-s S^{(2)}_{\Phi\Phi}}\right],
\end{equation}
where $S^{(2)}_{\Phi\Phi}$ denotes the scalar block of the Hessian, $\mathbf{Z}$ the corresponding matrix of kinetic prefactors, and $\mathrm{Tr}$ includes the functional trace and the internal $O(N)$ trace.

The scalar Hessian can be written as (see Ref.~\cite{Bonanno:2025qsc})
\begin{equation}
S^{(2)}_{\Phi\Phi}
=\int d^dx\sqrt g\;\frac{1}{2}\,\Phi^a\,H_{ab}(\rho)\,\Phi^b,
\qquad H_{ab}(\rho)=-Z_{ab}\,\Box+E_{ab}\,,
\end{equation}
with
\begin{equation}
Z_{ab}=\left\{
\begin{matrix}
Z_T(\Lambda) & a=b\neq N\\[0.3em]
Z_L(\Lambda)+4\bar\rho\,\frac{d-1}{d-2}\,\dfrac{\left[F'(\bar\rho)\right]^2}{F(\bar\rho)} & a=b=N\\[0.6em]
0 & a\neq b
\end{matrix}\right.
\end{equation}
and
\begin{equation}
\label{Eab}
E_{ab}=\Biggl(2\bar\rho\,U''(\bar\rho)+U'(\bar\rho)-\Biggl[2\bar\rho\,F''(\bar\rho)+F'(\bar\rho)-\frac{4\bar\rho}{d-2}\frac{\left[F'(\bar\rho)\right]^2}{F(\bar\rho)}\Biggr]\bar R\Biggr)P^{L}_{ab}+\Bigl(U'(\bar\rho)-F'(\bar\rho)\bar R\Bigr)P^{T}_{ab}\,.
\end{equation}
Here a bar denotes evaluation on the background, and $P^L_{ab}$ and $P^T_{ab}$ are the longitudinal (radial) and transverse projectors defined in the main text.

It is convenient to introduce the longitudinal and transverse ``mass'' terms
\begin{equation}
E_{ab}=M_L(\bar\rho)\,P^{L}_{ab}+M_T(\bar\rho)\,P^{T}_{ab}\,,
\end{equation}
with
\begin{equation}
\begin{split}
&M_L(\bar\rho)=2\bar\rho\,U''(\bar\rho)+U'(\bar\rho)-\Biggl[2\bar\rho\,F''(\bar\rho)+F'(\bar\rho)-\frac{4\bar\rho}{d-2}\frac{\left[F'(\bar\rho)\right]^2}{F(\bar\rho)}\Biggr]\bar R,\\
&M_T(\bar\rho)=U'(\bar\rho)-F'(\bar\rho)\bar R.
\end{split}
\end{equation}

\subsection{Heat-kernel expansion}
We bring the operator to Laplace type by factoring out $\mathbf{Z}$,
\begin{equation}
-\mathbf{Z}\Box+\mathbf{E}=\mathbf{Z}\left(-\Box+\mathbf{U}\right),\qquad \mathbf{U}\equiv\mathbf{Z}^{-1}\mathbf{E}.
\end{equation}
In components, $\mathbf{U}$ decomposes as
\begin{equation}
\mathbf{U}_{ab}=U_L\,P^L_{ab}+U_T\,P^T_{ab}, \qquad U_L=\frac{M_L}{Z_{NN}},\qquad U_T=\frac{M_T}{Z_{11}}=\frac{M_T}{Z_T(\Lambda)}.
\end{equation}
the term $U_L$ and $U_T$ correspond to the physical longitudinal and Goldstone mass mode, respectively.

For later reference, in a flat background ($\bar R=0$) and evaluated at the running minimum $U'(\bar\rho)=0$, the transverse mass $U_T$ vanishes (Goldstone modes), while the longitudinal one $U_L$ reduces to the mass term used in Eq.~(\ref{massL}) after conversion to dimensionless variables. In what follows we set $\bar R=0$, since $Z_L$ and $Z_T$ are extracted from the scalar kinetic terms and we focus on the flat-background projection.

The functional trace is evaluated using the early-time heat-kernel expansion for a Laplace-type operator
$\Delta=-\Box+\mathbf{U}$,
\begin{equation}
\label{HKexp}
\mathrm{Tr}\bigl[W(\Delta)\bigr]=\frac{1}{(4\pi)^{d/2}}\int d^dx\sqrt g\;\sum_{n=0}^{\infty} b_{2n}(\Delta)\,Q_{d/2-n}[W],
\end{equation}
with (see e.g.~\cite{Barvinsky:1990up,Barvinsky1987BeyondTS,Barvinsky:1990uq,Barvinsky:1993en,Avramidi:2000bm})
\begin{equation}
\begin{split}
&b_0(\Delta)=\mathrm{tr}\,\mathbf{1},\quad \quad b_2(\Delta)=\frac{1}{6}\,R\,\mathrm{tr}\,\mathbf{1}-\mathrm{tr}\,\mathbf{U},\\
&b_4(\Delta)=-\frac{1}{6}\Box\,\mathrm{tr}\,\mathbf{U}+O(R^2),\quad \quad
b_6(\Delta)=\mathrm{tr}\left[\frac{1}{12}(\partial_\mu\mathbf{U})(\partial^\mu\mathbf{U})+\frac{1}{6}\mathbf{U}\Box\mathbf{U}\right]+O(R^3).
\end{split}
\end{equation}
Here $\mathrm{tr}$ denotes the internal $O(N)$ trace only. The $Q$-functionals are Mellin transforms,
\begin{equation}
\label{Qdef}
Q_{p}[W]=\frac{1}{\Gamma(p)}\int_0^\infty dz\,z^{p-1}\,W(z)\,,\qquad p>0.
\end{equation}
where in our case, from eq.(\ref{projZ}), we read $W(z)=e^{-s(Az+B)}$.

Since we work on a background without boundary (and ultimately set $\bar R=0$), total derivatives can be dropped. In particular, the $b_4$ contribution is a total derivative, and in $b_6$ we may use $\Box(\mathbf{U}^2)=2\bigl(\mathbf{U}\Box\mathbf{U}+(\partial\mathbf{U})^2\bigr)$ to trade the $\mathbf{U}\Box\mathbf{U}$ term for $(\partial\mathbf{U})^2$ up to total derivatives. Thus, for the kinetic projection it is sufficient to keep the $(\partial\mathbf{U})^2$ structure.

\subsection{Derivative structure and separation into longitudinal/transverse sectors}

We compute
\begin{equation}
\mathrm{tr}\bigl[(\partial_\mu\mathbf{U})(\partial^\mu\mathbf{U})\bigr]=\sum_{a,b=1}^{N}\partial_\mu U_{ab}\,\partial^\mu U_{ba}.
\end{equation}
Using $\mathbf{U}=U_L P^L+U_T P^T$ and that $U_{L,T}$ are functions of $\rho$, one finds
\begin{equation}
\begin{split}
\mathrm{tr}\bigl[(\partial_\mu\mathbf{U})(\partial^\mu\mathbf{U})\bigr]
&=(\partial_\mu\rho)(\partial^\mu\rho)\left[(U_T')^2\sum_{a,b}P^T_{ab}P^T_{ab}+(U_L')^2\sum_{a,b}P^L_{ab}P^L_{ab}\right]\\
&\quad +(U_L-U_T)^2\sum_{a,b}(\partial_\mu P^L_{ab})(\partial^\mu P^L_{ab})\\
&\quad +2(U_L-U_T)(\partial_\mu\rho)\sum_{a,b}\left[P^L_{ab}(\partial^\mu P^L_{ab})U_L'+P^T_{ab}(\partial^\mu P^L_{ab})U_T'\right].
\end{split}
\end{equation}
The required projector identities are
\begin{equation}
\sum_{a,b}P^L_{ab}P^L_{ab}=1,\qquad \sum_{a,b}P^T_{ab}P^T_{ab}=N-1,\qquad \sum_{a,b}P^L_{ab}(\partial^\mu P^L_{ab})=0,\qquad \sum_{a,b}P^T_{ab}(\partial^\mu P^L_{ab})=0,
\end{equation}
and
\begin{equation}
\sum_{a,b}(\partial^\mu P^L_{ab})(\partial_\mu P^L_{ab})=-\frac{(\partial_\mu\rho)(\partial^\mu\rho)}{2\rho^2}+\frac{1}{\rho}\sum_{a=1}^{N}(\partial_\mu\phi_a)(\partial^\mu\phi_a).
\end{equation}
Therefore,
\begin{equation}\begin{split}
\label{dU2final}
&\mathrm{tr}\bigl[(\partial_\mu\mathbf{U})(\partial^\mu\mathbf{U})\bigr]=(\partial_\mu\rho)(\partial^\mu\rho)\left[(U_T')^2(N-1)+(U_L')^2\right]+\\
&\hspace{3cm}+(U_L-U_T)^2\left[-\frac{(\partial_\mu\rho)(\partial^\mu\rho)}{2\rho^2}+\frac{1}{\rho}\sum_{a=1}^{N}(\partial_\mu\phi_a)(\partial^\mu\phi_a)\right].
\end{split}\end{equation}

We now evaluate on an $O(N)$-breaking background pointing in the $N$-direction, so that only the longitudinal background component is nonzero. In this case $(\partial_\mu\rho)(\partial^\mu\rho)$ projects onto the longitudinal fluctuation sector, while the transverse sector is isolated by the term proportional to $\sum_{a=1}^{N-1}(\partial_\mu\phi_a)(\partial^\mu\phi_a)$. Explicitly, after evaluating on the background,
\begin{equation}
\label{b6split}
b_6(\Delta)
=\frac{1}{12}\,\mathrm{tr}\bigl[(\partial_\mu\mathbf{U})(\partial^\mu\mathbf{U})\bigr]=b_{6,L}\,(\partial_\mu\phi_N)(\partial^\mu\phi_N)+b_{6,T}\sum_{a=1}^{N-1}(\partial_\mu\phi_a)(\partial^\mu\phi_a),
\end{equation}
with
\begin{equation}
b_{6,L}=\frac{1}{12}\,2\bar\rho\left[(U_T')^2(N-1)+(U_L')^2\right], \qquad b_{6,T}=\frac{1}{12}\,(U_L-U_T)^2\frac{1}{\bar\rho}.
\end{equation}

\subsection{Proper-time integrals and flows of \texorpdfstring{$Z_L$}{ZL} and \texorpdfstring{$Z_T$}{ZT}}
Keeping only the $b_6$ term in Eq.~(\ref{HKexp}) and inserting Eq.~(\ref{b6split}) into the flow, we obtain
\begin{equation}
\begin{split}
\Lambda\partial_\Lambda S_\Lambda
&=\frac{1}{2}\int_{0}^{\infty}\frac{ds}{s}\,r\!\left(s,\Lambda^2\mathbf{Z}\right)\, \frac{1}{(4\pi)^{d/2}}\int d^dx\sqrt g\;b_6(\Delta)\,Q_{d/2-3}[W]\\
&=\frac{1}{(4\pi)^{d/2}}\int d^dx\sqrt g\Bigg\{\left[\frac{1}{2}\int_{0}^{\infty}\frac{ds}{s}\,r\!\left(s,\Lambda^2 Z_T\right)\,b_{6,T}\,Q_{d/2-3}\!\left(W_T\right)\right]\sum_{a=1}^{N-1}(\partial_\mu\phi_a)(\partial^\mu\phi_a)\\
&\hspace{8.8em}+\left[\frac{1}{2}\int_{0}^{\infty}\frac{ds}{s}\,r\!\left(s,\Lambda^2 Z_L\right)\,b_{6,L}\,Q_{d/2-3}\!\left(W_L\right)\right]
(\partial_\mu\phi_N)(\partial^\mu\phi_N)\Bigg\},
\end{split}
\end{equation}
where
\begin{equation}
W_L(z)=\exp\!\left[-s\left(Z_{NN}z+M_L\right)\right],
\qquad
W_T(z)=\exp\!\left[-s\left(Z_T z+M_T\right)\right].
\end{equation}
Matching to the left-hand side of Eq.~(\ref{projZ}) yields
\begin{equation}
\begin{split}
&\Lambda\partial_\Lambda Z_L=-\frac{1}{(4\pi)^{d/2}}\int_{0}^{\infty}\frac{ds}{s}\,r\!\left(s,\Lambda^2 Z_L\right)\; \frac{1}{12}\,2\bar\rho\left[(U_T')^2(N-1)+(U_L')^2\right]\, Q_{d/2-3}\!\left(W_L\right),\\
&\Lambda\partial_\Lambda Z_T=-\frac{1}{(4\pi)^{d/2}}\int_{0}^{\infty}\frac{ds}{s}\,r\!\left(s,\Lambda^2 Z_T\right)\; \frac{1}{12}\,(U_L-U_T)^2\frac{1}{\bar\rho}\, Q_{d/2-3}\!\left(W_T\right).
\end{split}
\end{equation}

Using the cutoff kernel in Eq.~(\ref{cutoff}) (type-C: $\epsilon=0$) one may perform the $z$-integral in $Q_p$ and then the $s$-integral. A convenient identity is
\begin{equation}
\label{PTidentity}
\frac{1}{2}\int_{0}^{\infty}\frac{ds}{s}\,r\!\left(s,\Lambda^2 A\right)\, Q_{p}\!\left(W=e^{-s(Az+B)}\right)=\left(\gamma\Lambda^2\right)^{p}\left(1+\frac{B}{A\gamma\Lambda^2}\right)^{p-m}\, \frac{\Gamma(m-p)}{(4\pi)^{d/2}\Gamma(m)},\qquad p=\frac{d}{2}-j,
\end{equation}
which is to be used with $A=Z_{NN}$, $B=M_L$ (longitudinal) or $A=Z_T$, $B=M_T$ (transverse), and $j=3$ for the present projection.

Inserting Eq.~(\ref{PTidentity}) and rewriting $\partial_t\equiv \Lambda\partial_\Lambda$ finally yields
\begin{equation}
\begin{split}
&\partial_t Z_L=-\frac{\left(\gamma \Lambda^2\right)^\frac{d-6}{2}}{3\left(4\pi\right)^\frac{d}{2}}\frac{\Gamma\left(m-\frac{d}{2}+3\right)}{\Gamma\left(m\right)}\left(1+\frac{{\bar{U}}^\prime+2\bar{\rho}{\bar{U}}^{\prime\prime}}{\gamma \Lambda^2Z_L\left(1+\frac{4\left(d-1\right)\bar{\rho}\left({\bar{F}}^\prime\right)^2}{\left(d-2\right)Z_L\bar{F}}\right)}\right)^{\frac{d}{2}-\left(m+3\right)}\Bigg[\frac{\bar{\rho}\left({\bar{U}}^{\prime\prime}\right)^2}{Z_T^2}\left(N-1\right)+\\
&+\frac{4{\bar{\rho}}^2{\bar{U}}^{\prime\prime\prime}-3{\bar{U}}^\prime}{4\bar{\rho}Z_L^2\left(1+\frac{4\left(d-1\right)\bar{\rho}\left({\bar{F}}^\prime\right)^2}{\left(d-2\right)Z_L\bar{F}}\right)^2}\left(1+\frac{{\bar{U}}^\prime+2\bar{\rho}{\bar{U}}^{\prime\prime}}{4{\bar{\rho}}^2{\bar{U}}^{\prime\prime\prime}-3{\bar{U}}^\prime}\left(1+\frac{2\left(1+\frac{4\left(d-1\right){\bar{\rho}}^2{\bar{F}}^\prime}{\left(d-2\right)Z_L{\bar{F}}}\left(\frac{\left({\bar{F}}^\prime\right)^2}{{\bar{F}}}-2{\bar{F}}^{\prime\prime}\right)\right)}{1+\frac{4\left(d-1\right)\bar{\rho}\left({\bar{F}}^\prime\right)^2}{\left(d-2\right)Z_L\bar{F}}}\right)\right)^2\Bigg],
\end{split}
\end{equation}
and
\begin{equation}
\begin{split}
\partial_t Z_T
&=-\frac{\left(\gamma \Lambda^2\right)^{\frac{d-6}{2}}}{6(4\pi)^{d/2}\bar\rho} \frac{\Gamma\!\left(m-\frac{d}{2}+3\right)}{\Gamma(m)} \left(1+\frac{{\bar{U}}'}{\gamma \Lambda^2 Z_T}\right)^{\frac{d}{2}-(m+3)} \left(\frac{{\bar{U}}'}{Z_T}-\frac{{\bar{U}}'+2\bar\rho{\bar{U}}''}{Z_L\left(1+\frac{4(d-1)\bar\rho({\bar{F}}')^2}{(d-2)Z_L{\bar{F}}}\right)}\right)^2.
\end{split}
\end{equation}
A bar indicates evaluation on the background. Using the definitions in Eq.~(\ref{defeta}) and the dimensionless variables $x=Z_L\Lambda^{2-d}\rho$, $u_\Lambda(x)=\Lambda^{-d}U_\Lambda(\rho)$, and $f_\Lambda(x)=\Lambda^{2-d}F_\Lambda(\rho)$, one obtains the anomalous dimensions reported in Eq.~(\ref{etaLetaT}).

\subsection{Consistency check: the \texorpdfstring{$N=1$}{N=1} limit}
It is instructive to verify how the $N=1$ case is recovered. In Eq.~(\ref{dU2final}), the transverse contribution is proportional to
\[
-\frac{(\partial_\mu\rho)(\partial^\mu\rho)}{2\rho^2}+\frac{1}{\rho}\sum_{a=1}^{N}(\partial_\mu\phi_a)(\partial^\mu\phi_a).
\]
For $N=1$, $\rho=\phi^2/2$ and $(\partial_\mu\rho)(\partial^\mu\rho)=\phi^2(\partial_\mu\phi)(\partial^\mu\phi)$, hence
\begin{equation}
-\frac{(\partial_\mu\rho)(\partial^\mu\rho)}{2\rho^2}+\frac{1}{\rho}(\partial_\mu\phi)(\partial^\mu\phi)=-\frac{\phi^2(\partial_\mu\phi)(\partial^\mu\phi)}{2(\phi^4/4)}+\frac{(\partial_\mu\phi)(\partial^\mu\phi)}{\phi^2/2}=0.
\end{equation}
Therefore only the longitudinal contribution remains, as expected.

\section{The beta functions}\label{appbeta}
\noindent In this appendix we derive the beta functions for the ansats 
\begin{equation}\begin{split}
\label{ansappbeta}
&u_t(x)=u_0(t)+\lambda(t)\left(x-x_0(t)\right)^2+O\left(x-x_0(t)\right)^3\\
&f_t(x)=\frac{1}{g(t)}+f_1(t)\left(x-x_0(t)\right)+O\left(x-x_0(t)\right)^2.
\end{split}\end{equation}
The beta functions can be found replacing the ansats in eqs.(\ref{adeqfull}) and expanding in series around $x=x_0$. 

In $d=4$ the flow equation for $u_0$ is given by
\begin{equation}
{\dot{u}}_0=-4u_0+\frac{\gamma^2}{16\pi^2}\frac{\Gamma\left(m-2\right)}{\Gamma\left(m\right)}\left(N+1+\left(1+\frac{4\lambda x_0}{\gamma\left(1+6f_1^2gx_0\right)}\right)^{2-m}\right)
\end{equation}
The flow equation for $x_0$ is given by
\begin{equation}
{\dot{x}}_0=-\left(2+\eta_L\right)x_0+\frac{\gamma}{16\pi^2\left(m-1\right)}\left(\frac{N-1}{w}+\frac{3\left(1+2f_1^2gx_0+4f_1^3g^2x_0^2\right)}{\left(1+6f_1^2gx_0\right)^2\left(1+\frac{4\lambda x_0}{\gamma\left(1+6f_1^2gx_0\right)}\right)^{m-1}}\right)
\end{equation}
The flow equation for $\lambda$ is given by
\begin{equation}\begin{split}
&\dot{\lambda}=\lambda\left(2\eta_L+\frac{9\gamma f_1^2g\left(1-f_1gx_0\right)\left(1+2f_1gx_0\left(f_1-\frac{1}{3}\right)\right)}{4\pi^2\left(m-1\right)\left(1+6f_1^2gx_0\right)^3\left(1+\frac{4\lambda x_0}{\gamma\left(1+6f_1^2gx_0\right)}\right)^m}\right)+\\
&+\frac{\lambda^2}{8\pi^2\left(m-1\right)}\left(\frac{\left(m-1\right)\left(N-1\right)}{w^2}+\frac{3C\left(t\right)}{\left(1+6f_1^2gx_0\right)^4\left(1+\frac{4\lambda x_0}{\gamma\left(1+6f_1^2gx_0\right)}\right)^m}\right)
\end{split}\end{equation}
where
\begin{equation}\begin{split}
&C\left(t\right)=3\left(m-1\right)+12g\left(1+m\right)f_1^2x_0+4f_1^3g^2x_0^2\left(6m-16+3\left(3+m\right)f_1\right)+\\
&+16f_1^4g^3x_0^3\left(1+3\left(m-2\right)f_1\right)+48f_1^6g^4x_0^4\left(m-1\right)
\end{split}\end{equation}
The flow equation for $g$ is given by
\begin{equation}
\dot{g}=2g+\frac{\gamma g^2}{6\pi^2\left(m-1\right)}\left(\frac{N-1}{16}-1+\frac{1-12f_1+12f_1^2gx_0\left(2+3gf_1^2x_0\right)}{16\left(1+6f_1^2gx_0\right)^2\left(1+\frac{4\lambda x_0}{\gamma\left(1+6f_1^2gx_0\right)}\right)^{m-1}}\right)
\end{equation}
The flow equation of $f_1$ is given by
\begin{equation}\begin{split}
&{\dot{f}}_1=f_1\eta_L+\frac{1}{8\pi^2\left(m-1\right)}\Bigg\{\frac{\lambda}{w}\left(\frac{1}{6}+\frac{f_1}{w}\right)\left(m-1\right)\left(N-1\right)+\\
&+\frac{1}{\left(1+6f_1^2gx_0\right)^2\left(1+\frac{4\lambda x_0}{\gamma\left(6f_1^2gx_0+1\right)}\right)^m}\Bigg[3\gamma f_1^2g\left(f_1-\frac{1}{3}\right)\left(1-f_1gx_0\right)\left(1+\frac{4\lambda x_0}{\gamma\left(6f_1^2gx_0+1\right)}\right)+\\
&+\frac{\lambda\left(m-1\right)\left(1+2f_1^2gx_0\left(1+2f_1gx_0\right)\right)\left(1+6f_1\left(1+3f_1gx_0\right)\right)}{2\left(1+6f_1^2gx_0\right)}\Bigg]\Bigg\}
\end{split}\end{equation}
All beta functions do not depend on $u_0$, consequently the equations are decoupled with respect to the flow of $u_0$. In the flow equations, the quantity $\frac{4\lambda x_0}{1+6f_1^2gx_0}$ is the dimensionless longitudinal scalar mass. The flow equations do not depend on $\eta_T$ and the presence of $Z_T$ appears only by $w$. The presence of $\eta_L$ breaks the classical marginality of $\lambda$, $f_1$ and shifts the classical scaling of $x_0$. The flow of $g$ can be written in the standard form $\dot g=(d-2+\eta_g)g$. The gravitational anomalous dimension $\eta_g$ does not depend on $w$ and $\eta_L$ whereas $\lambda$ appears only by the presence of the longitudinal scalar mass. 

In the full flow equations eqs.(\ref{adeqfull}) the potential is coupled to gravity only by $f'(x)$. In term of ansats eq.(\ref{ansappbeta}), this translates in the flow of $\lambda$  coupled to the newtonian constant only by $f_1$. This is a feature of the physical gauge. The origin of this can be found writing the pysical gauge in terms of background field gauge method. With the standard notation of \cite{Ohta:2021bkc}, the physical gauge is the limit $\beta\to\infty$ and $\alpha\to0$. In the flow equations for general $\beta$ and $\alpha$ the newtonian constant always appears in the form $\sim g/\beta$ \cite{Ohta:2021bkc}. Accordingly, in the limit $\beta\to\infty$ the standard terms disappear and $g$ is coupled to the potential only if the newtonian constant has a field dependence.

The flow equation for $w(t)$ is given by 
\begin{equation}
\dot w(t)=w(t)(\eta_L-\eta_T)
\end{equation}
The equation for $\eta_L$ and $\eta_T$ can be derived simply replacing the ansatz in eq.(\ref{etaLetaT}) and putting $x=x_0$. The results are
\begin{equation}\begin{split}
\label{betaetaL}
&\eta_L=\frac{\lambda^2x_0m}{12\pi^2\gamma\left(1+\frac{4\lambda x_0}{\gamma\left(1+6f_1^2gx_0\right)}\right)^{m+1}}\left(\frac{N-1}{w^2}+9\left(\frac{1+2f_1^2gx_0\left(1+2f_1gx_0\right)}{\left(1+6f_1^2gx_0\right)^2}\right)^2\right)\\
&\eta_T=\frac{\lambda^2mx_0}{6\gamma w\pi^2\left(1+6f_1^2gx_0\right)^2}
\end{split}\end{equation}
In this way the flow equation of $w$ is given by
\begin{equation}
\dot{w}=\frac{\lambda^2mx_0}{6\gamma\pi^2}\left[\frac{\frac{N-1}{w}+9w\left(\frac{1+2f_1^2gx_0\left(1+2f_1gx_0\right)}{\left(1+6f_1^2gx_0\right)^2}\right)^2}{2\left(1+\frac{4\lambda x_0}{\gamma\left(6f_1^2gx_0+1\right)}\right)^{m+1}}-\frac{1}{\left(1+6f_1^2gx_0\right)^2}\right]
\end{equation}
from eq.(\ref{betaetaL}) we see that at the fixed point $\lambda_*=0$ the two anomalous dimensions vanish, so $w_*$ is a free parameter.

\section{Perturbations of the scaling solution \texorpdfstring{$u=u_*$}{N} and \texorpdfstring{$f=f_*x$}{N}}\label{apppert}
\noindent In this appendix, we derive the perturbations to the exact scaling solution $u=u_*$ and $f=f_{1*}x$ of eq.(\ref{adeqfull}), which have not been discussed in \cite{Bonanno:2025qsc}. The following discussion shed light on properties of the  UV complete solutions we discussed in the main text but also give a more general insight far from the minimum of the potential.

The values of the coefficients $u_*$ and $f_{*1}$ of the exact scaling solution are given in eq.(\ref{FPbello}). We consider the perturbations of the solution with the plus sign, which yields the UV complete solutions discussed in the text.

Before proceeding, we emphasize that the scaling solution $u = u_*$, $f = f_{1*} x$ is defined only for $x \neq 0$. At $x = 0$, the gravitational coupling vanishes, which invalidates the ansatz $f = f_{1*} x$. In fact, the configuration $u = u_*$, $f = 0$ does not satisfy the fixed-point version of Eq.~(\ref{adeqfull})\footnote{The fixed-point version of Eq.~(\ref{adeqfull}) with the ansatz $u = u_*$ and $f = 0$ is satisfied only for $N = 16$. However, this value is pathological, as it corresponds to a negative Newton constant.}. This implies that a polynomial expansion around $x = 0$ for the solutions of Eq.~(\ref{adeqfull}) does not admit a fixed point that reconstructs $f = f_{1*} x$ in the UV limit ($t \to +\infty$). This is in contrast to the expansion around $x = x_0$, where Eq.~(\ref{FPbello}) yields $1/g_* = f_{1*} x_{0*}$, which is equivalent to the lower bound of the critical condition in Eq.~(\ref{critval}). As a consequence, in the symmetric phase there is no UV-complete solution exhibiting the scaling regime $u = u_*$, $f = f_{1*} x$. In particular, the removal of the Landau pole in the symmetric phase cannot be achieved within a simple truncation of the form $F(x) R + U(x)$, as Eq.~(\ref{adeqfull}) does not admit a scaling solution capable of realizing this behavior. In this case, additional fields, interactions, or extensions of the derivative expansion are required.

The general perturbations of the scaling solution can be written as
\begin{equation}
u(x,t)=u_*+\varepsilon\delta u(x,t), \quad f(x,t)=f_*x+\varepsilon \delta f(x,t) , \quad w(t)=w_*+\varepsilon \delta w(t)
\end{equation}
replacing them in the flow equations and expanding around $\varepsilon=0$ at the linear order we get
\begin{equation}\begin{split}
\label{eqpertuf}
&\delta \dot u=-4\delta u-\left(B_u-2x\right){\delta u}^\prime-C_ux{\delta u}^{\prime\prime}\\
&\delta \dot f=-\left(2+\frac{A_f}{x}\right)\delta f-\left(B_f-2x\right)\delta f^\prime-C_fx\delta f^{\prime\prime}-D_{uf}\delta u^\prime-E_{uf}x\delta u^{\prime\prime}-G_{wf}\delta w\\
&\delta \dot w =0
\end{split}\end{equation}
where
\begin{equation}\begin{split}
&B_u=\frac{\gamma}{96\pi^2\left(m-1\right)}\left(N-13+\frac{5\left(N-1\right)}{w_\ast}+\sqrt{\left(N-13\right)^2-\frac{2\left(N-19\right)\left(N-1\right)}{w_\ast}-\frac{\left(N-1\right)^2}{w_\ast^2}}\right)\\
&C_u=\frac{\gamma\left(N-1\right)}{4\pi^2\left(m-1\right)w_\ast\left(-\left(N-13\right)+\frac{N-1}{w_\ast}+\sqrt{\left(N-13\right)^2-\frac{2\left(N-19\right)\left(N-1\right)}{w_\ast}+\frac{\left(N-1\right)^2}{w_\ast^2}}\right)}
\end{split}\end{equation}
and
\begin{equation}\begin{split}
&A_f=-\frac{\gamma\left(N-1\right)\left(N-13+\frac{5\left(N-1\right)}{w_\ast}+\sqrt{\left(N-13\right)^2-\frac{2\left(N-19\right)\left(N-1\right)}{w_\ast}+\frac{\left(N-1\right)^2}{w_\ast^2}}\right)}{8\pi^2\left(m-1\right)\left(-\left(N-13\right)+\frac{N-1}{w_\ast}+\sqrt{\left(N-13\right)^2-\frac{2\left(N-19\right)\left(N-1\right)}{w_\ast}+\frac{\left(N-1\right)^2}{w_\ast^2}}\right)^2}\\
&B_f=\frac{\gamma}{96\pi^2\left(m-1\right)}\Bigg[\left(N-13\right)\left(2N-27\right)-\frac{\left(4N-71\right)\left(N-1\right)}{w_\ast}+\frac{2\left(N-1\right)^2}{w_\ast^2}+\\
&+\left(2N-27-\frac{2\left(N-1\right)}{w_\ast}\right)\sqrt{\left(N-13\right)^2-\frac{2\left(N-19\right)\left(N-1\right)}{w_\ast}+\frac{\left(N-1\right)^2}{w_\ast^2}}\Bigg]\\
&C_f=C_u
\end{split}\end{equation}
The mixing coefficients are given by
\begin{equation}\begin{split}
&D_{uf}=\frac{1}{576\pi^2}\Bigg[\left(N-17\right)\left(N-13\right)+\frac{\left(35-2N\right)N-69}{w_\ast}+\frac{\left(N-1\right)\left(N+2\right)}{w_\ast^2}+\\
&+\left(N-17-\frac{\left(N+2\right)}{w_\ast}\right)\sqrt{\left(N-13\right)^2-\frac{2\left(N-19\right)\left(N-1\right)}{w_\ast}+\frac{\left(N-1\right)^2}{w_\ast^2}}\Bigg]\\
&E_{uf}=\frac{\left(N-1\right)\left(2\left(N-13\right)+\frac{N-1}{w_\ast}-2\sqrt{\left(N-13\right)^2-\frac{2\left(N-19\right)\left(N-1\right)}{w_\ast}+\frac{\left(N-1\right)^2}{w_\ast^2}}\right)}{12\pi^2\left(-\left(N-13\right)+\frac{N-1}{w_\ast}+\sqrt{\left(N-13\right)^2-\frac{2\left(N-19\right)\left(N-1\right)}{w_\ast}+\frac{\left(N-1\right)^2}{w_\ast^2}}\right)^2}\\
&G_{wf}=\frac{\gamma}{192\pi^2\left(m-1\right)}\left(\frac{N-13}{w_\ast}+\frac{N-1}{w_\ast^2}-\frac{1}{w_\ast}\sqrt{\left(N-13\right)^2-\frac{2\left(N-19\right)\left(N-1\right)}{w_\ast}+\frac{\left(N-1\right)^2}{w_\ast^2}}\right)
\end{split}\end{equation}
The perturbations for $\delta u$ and $\delta f$ are two second order linear partial differential equations. In particular, the equation for $\delta u$ is a homogeneous equation $\mathcal{L}_u\delta u=0$ decoupled from $\delta f$, whereas the equation for $\delta f$ is a non-homogeneous equation   $\mathcal{L}_f\delta f=T(x,t)$. The perturbation for $\delta w$ is a free parameter. This is due to the perturbations of the anomalous dimensions, which are always of the second order in the perturbative expansion.

The equation for $\delta u$ can be solved exactly. A solution can be found with the method of the separation of variables with ansats $\delta u=q_u(x)e^{-\theta_u t}$ where $\theta_u$ is the separation constant, which describes the critical exponents of the solution. The function $q_u(x)$ then satisfies a linear second order differential equation whose solution is given by
\begin{equation}
\label{solpertu}
q_u\left(x\right)=c_{1u}U{\left(\frac{\theta_u}{2}-1,\frac{B_u}{C_u},\frac{2x}{C_u}\right)}+c_{2u}L_{-\frac{\theta_u}{2}+1}^{\frac{B_u}{C_u}-1}\left(\frac{2x}{C_u}\right)
\end{equation}
where $U\left(a,b,c\right)$ is the Tricomi confluent hypergeometric function, $L_n^a\left(x\right)$ are the generalized Laguerre polynomial and $c_{1u}$, $c_{2u}$ are two free parameters. The general solution is a superposition of $\delta u=q_u(x)e^{-\theta_u t}$.

The equation for $\delta f$ has a general solution that can always be written as
\begin{equation}
\label{solfpert}
\delta f(x,t)=\delta f_{hom}(x,t)+\delta f_{part}(x,t)
\end{equation}
where $\delta f_{hom}(x,t)$ solves the homogeneous equation $\mathcal{L}_f\delta f=0$ and $\delta f_{part}(x,t)$ is a particular solution. The homogeneous solution can be found again with the method of separation of variables with an ansats $\delta f_{hom}=q_f(x)e^{-\theta_f t}$. The separation constant $\theta_f$  does not necessarily coincide with $\theta_u$ as the differential equations of $\delta u$ and $\delta f$ are decoupled. The $x$ dependent part of the homogeneous solution is given by 
\begin{equation}\begin{split}
\label{solpertf}
&q_f\left(x\right)=x^{\frac{1}{2}-\frac{B_f}{2C_f}+\frac{1}{2}\alpha}\left[c_{1f}U{\left(\frac{\theta_f-1}{2}-\frac{B_f}{2C_f}+\frac{1}{2}\alpha,1+\alpha,\frac{2x}{C_f}\right)}
+c_{2f}L_{\frac{1-\theta_f}{2}+\frac{B_f}{2C_f}-\frac{1}{2}\alpha}^{\alpha}\left(\frac{2x}{C_f}\right)\right]
\end{split}\end{equation}
where
\begin{equation}
\alpha=\sqrt{1+\frac{B_f^2}{C_f^2}-\frac{2B_f}{C_f}-\frac{4A_f}{C_f}}
\end{equation}
The particular solution is given formally by
\begin{equation}
\delta f_{part}(x,t)=\mathcal{L}_f^{-1}T(x,t)=\int{G(x,x',t)T(x',t)dx'}
\end{equation}
where $G(x,x',t)$ is the Green function of the differential operator $\mathcal{L}_f$. Due to the complexity of the Green function, the particular solution can be found only numerically, except in the asymptotics or locally around some point $x=\bar x$. 

The critical properties of the theory depend only on the homogeneous solution. In order to obtain well-defined critical exponents, we have to require that both branches of the solution are regular over the entire range of $x$, from zero to infinity. While the Laguerre polynomials are smooth functions, the Tricomi confluent hypergeometric function $U(a,b,c)$, with argument $c = px$, exhibits a singular behavior around $x = 0$:
\begin{equation}
U\left(a,b,px\to0\right)=x^{-b}\left(\frac{p^{1-b}\Gamma\left(b-1\right)}{\Gamma\left(a\right)}x+O(x^2)\right)+\left(\frac{\Gamma\left(1-b\right)}{\Gamma\left(1+a-b\right)}+O{\left(x^1\right)}\right)
\end{equation}
If $b > 1$, this leads to a singularity at $x = 0$. In order to avoid this singular behavior, $\Gamma(a)$ must also be singular, which occurs when $a = -j$, with $j \in \mathbb{N}$.

Beyond regularity, one must also require that the solution does not exhibit exponential growth at large $x$. In the asymptotic limit $x \to \infty$, the Tricomi function yields a power-law behavior, whereas the generalized Laguerre polynomial $L_a^b(px)$ behaves as
\begin{equation}
L_a^b(px \to \infty) = \frac{e^{px}}{x^{a+b}} \left( \frac{1}{p^{1+a+b}} \frac{\Gamma(a+b+1)}{\Gamma(a+1)\Gamma(-a)} \frac{1}{x} + O\!\left(\frac{1}{x^2}\right) \right)
+ x^a \left( \frac{(-p)^a}{\Gamma(a+1)} + O\!\left(\frac{1}{x}\right) \right).
\end{equation}
Therefore, in order to eliminate the exponential term, the condition $a+1 = -j$ or $-a = -j$ must be satisfied. 

In Eqs.~(\ref{solpertu}) and (\ref{solpertf}), the parameter $a$ of the generalized Laguerre polynomial is equal in magnitude but opposite in sign to that of the Tricomi function. As a result, imposing the condition $a = -j$ simultaneously removes both the singularity at the origin and the exponential growth at infinity.

In our case, for fixed $N > 1$ and $w_\ast$, both $U(a,b,c)$ appearing in $q_u$ and $q_f$ have $b > 1$. Therefore, imposing $a = -j$ to remove the divergences at $x = 0$ and the exponential behavior at $x \to \infty$ uniquely fixes the critical exponents. In particular, this leads to two independent sets of critical exponents.

From Eq.~(\ref{solpertu}) we obtain
\begin{equation}
\label{critexpu}
\theta_u = 4 - 2 j_u, \qquad j_u = 0,1,\ldots
\end{equation}
This result does not depend on $N$ or on the regulator parameter $m$. The values $j_u = 0$ and $j_u = 1$ correspond to relevant directions, $j_u = 2$ yields a marginal direction, and $j_u > 2$ correspond to irrelevant directions. This explains the origin of the integer critical exponents $\theta = 4, 2, 0$ associated with the UV-complete solutions discussed in Sec.~\ref{sectionfixedpoint}.

More generally, repeating the analysis of Sec.~\ref{sectionfixedpoint} with a truncation of the form
\[
u(x,t) = u_0(t) + \lambda(t)(x - x_0(t))^2 + \sum_{n=3}^\infty u_n(t)(x - x_0(t))^n
\]
and comparing with Eq.~(\ref{critexpu}) shows that the renormalizable couplings $u_0$ and $\lambda$ are associated with the directions $\theta_u = 4$ and $\theta_u = 0$, respectively, while the higher-order couplings $u_3$, $u_4$, \dots correspond to irrelevant directions. The value $\theta_u = 2$ is instead associated with the scalar mass generated by the non-trivial potential.

Replacing eq.(\ref{critexpu}) in the pertubation $q_u(x)$, eq.(\ref{solpertu}), turns the solution into a simple polynomial of degree $j_{u}$. The general solution for $\delta u$ is given by
\begin{equation}
\label{soluvinc}
\delta u(x,t)=\sum_{j_{u}=0}^{+\infty}q_u\left(x,j_u\right)e^{-\theta_{j_{u}}t}
\end{equation}
where
\begin{equation}
\label{solqu}
q_u\left(x,j_u\right)=\sum_{n=0}^{j_u}{\left(j_u !c_{1u}+\left(-1\right)^{j_u}c_{2u}\right)a_n\left(m,N,w_\ast\right)x^n}
\end{equation}
the coefficients $a_n(m,N,w_*)$ are complicated and long functions of $N$ and $w_\ast$, which we do not show explicitly.  

From eq.(\ref{solpertf}) we get
\begin{equation}\begin{split}
\label{critexpf}
&\theta_f=-2j_f+\frac{N-11}{4}-\frac{N-1}{4w_\ast}+\frac{3}{4}\sqrt{\left(N-13\right)^2-\frac{2\left(N-19\right)\left(N-1\right)}{w_\ast}+\frac{\left(N-1\right)^2}{w_\ast^2}}-\\
&-\frac{1}{4}\left(\frac{A_\theta}{-\left(N-13\right)+\frac{N-1}{w_\ast}+\sqrt{\left(N-13\right)^2-\frac{2\left(N-19\right)\left(N-1\right)}{w_\ast}+\frac{\left(N-1\right)^2}{w_\ast^2}}}\right)^\frac{1}{2}\ ,\ \ j_f=0,1,\ldots
\end{split}\end{equation}
where
\begin{equation}\begin{split}
&A_\theta=-\left(N-13\right)\left(\left(N-20\right)N+92\right)+\frac{\left(3\left(N-25\right)N+409\right)\left(N-1\right)}{w_\ast}-\frac{3\left(N-14\right)\left(N-1\right)^2}{w_\ast^2}+\frac{\left(N-1\right)^3}{w_\ast^3}+\\
&+\left(\left(N-20\right)N+92-\frac{\left(2N-47\right)\left(N-1\right)}{w_\ast}+\frac{\left(N-1\right)^2}{w_\ast^2}\right)\sqrt{\left(N-13\right)^2-\frac{2\left(N-19\right)\left(N-1\right)}{w_\ast}+\frac{\left(N-1\right)^2}{w_\ast^2}}
\end{split}\end{equation}
This result does not depend on the regulator parameter $m$, but it does depend on $N$ and $w_*$. The number of relevant directions is given by $n_{\text{rel}} = 1 + j_{\text{crit}}$, where $j_{\text{crit}}$ is defined as $j_{\text{crit}} = \left\lfloor \frac{Q}{2} \right\rfloor$, with $Q$ given in Eq.~(\ref{critexpf}). By plotting $Q$ as a function of $w_*$ for different values of $N$, we find $j_{\text{crit}} = 1$ for $N \leq 5$ and $j_{\text{crit}} = 0$ for $N > 5$. Consequently, $n_{\text{rel}} = 2$ for $N \leq 5$ and $n_{\text{rel}} = 1$ for $N > 5$. This matches the number of non-trivial critical exponents found in Sec.~\ref{sectionfixedpoint}. In particular, the values are very close to those shown in Figs.~\ref{plottheta2} and \ref{plottheta1}. For example, for $N = 4$ and $w_* = 1$, we obtain $\theta_f = 2.052$ ($j_f = 0$) and $\theta_f = 0.052$ ($j_f = 1$), while the corresponding values extracted from the plots are $\theta = 2.032$ and $\theta = 0.031$. The small discrepancy between these results is due to truncation effects. Repeating the analysis of Sec.~\ref{sectionfixedpoint} with higher-order truncations in the expansion of $f$ around $(x - x_0(t))$, the values $\theta = 2.032$ and $\theta = 0.031$ converge toward $\theta_f = 2.052$ and $\theta_f = 0.052$. This explains the origin of the two non-trivial critical exponents associated with the UV-complete solutions.

As for $\delta u$, replacing eq.(\ref{critexpf}) in the pertubation $q_f(x)$, eq.(\ref{solpertf}), turns the solution into a simple polynomial of degree $j_{f}$. The general homogeneous solution for $\delta f$ is given by
\begin{equation}
\delta f_{hom}(x,t)=\sum_{j_{f}=0}^{+\infty}q_f\left(x,j_f\right)e^{-\theta_{j_{f}}t}
\end{equation}
where
\begin{equation}
\label{solqf}
q_f(x,j)=x^{\frac{1}{2}-\frac{B_f}{2C_f}+\frac{1}{2}\sqrt{1+\frac{B_f^2}{C_f^2}-\frac{2B_f}{C_f}-\frac{4A_f}{C_f}}}\sum_{n=0}^{j_f}{\left(j_f !c_{1f}+\left(-1\right)^{j_f}c_{2f}\right)b_n\left(m,N,w_\ast\right)x^n}.
\end{equation}
the coefficients $b_n(m,N,w_*)$ are complicated and long functions of $N$ and $w_\ast$, which we do not show explicitly. 

Replacing eq.(\ref{soluvinc}) in the perturbation equation of $\delta f(x,t)$ and expanding around $x=0$ with a Frobenius ansats for $\delta f$, we find that the particular solution in this regime can be written as
\begin{equation}
\delta f_{part}(x\to0,t)=x\sum_{n=0}^{+\infty}{c_n(t) x^n}
\end{equation}
for all $j_u$, where the coefficients $c_n(t)$ are a linear combination of $e^{-\theta_u t}$ for $j_{u}\ge1$: 
\begin{equation}
c_n(t)=\sum_{j_u=1}^{+\infty}\alpha_n(N,w_*,m)e^{-\theta_{j_{u}}t}
\end{equation}
the coefficients $\alpha_n(N,w_*,m)$ are long and complicated functions of $N$, $w_*$ and $m$ that we do not show explicitly. The relevant direction $\theta=4$ does not appear because this yields a constant eigenfunction and in eq.(\ref{eqpertuf}) only derivatives of $\delta u$ appear. 

The particular solution for $\delta f$ is always a smooth function around the origin. A similar conclusion holds for an expansion around a generic $x=\bar x$ where we find 
\begin{equation}
\delta f_{part}(x\to\bar x,t)=\sum_{n=0}^{+\infty}{d_n(t)(x-\bar x)^n}
\end{equation}
where the coefficients $d_n(t)$ have the same structure as $c_n(t)$. A particular case arises when $\bar x$ coincides with the running minimum $x_0(t)$ of the potential. In this case, $\bar x$ acquires an explicit $t$-dependence and the resulting solution is modified. This situation will be analyzed in the following.

A different situation arises in the asymptotic regime $x \to \infty$, where the leading order of the Frobenius ansatz depends on the largest $\theta_{j_u}$ included in the expansion. Truncating the series in Eq.~(\ref{soluvinc}) at $j_u = j_{\max}$, the particular solution in this regime can be written as
\begin{equation}
\delta f_{\text{part}}(x \to \infty, t) = x^{j_{\max}-1} \sum_{i=0}^{+\infty} \frac{h_i(t)}{x^i} \, .
\end{equation}
Here,
\begin{equation}
h_i(t) = i \sum_{l=0}^{i-1} \alpha_{l,i}(N,w_*,m)\, e^{-(\theta_{j_{\max}+1} + 2l)t}
+ t \sum_{l=0}^{i} \beta_{l,i}(N,w_*,m)\, e^{-(\theta_{j_{\max}+1} + 2l)t} \, ,
\end{equation}
so that the coefficients $h_i(t)$ consist of superpositions of exponential terms $e^{-t \theta_{j_u}}$, together with contributions linear in $t$ multiplied by the same exponential factors.

Fixing $N > 1$ and $w_\ast$, the exponent of the overall $x$ coefficient
\[
\frac{1}{2} - \frac{B_f}{2C_f} + \frac{1}{2}\sqrt{1 + \frac{B_f^2}{C_f^2} - \frac{2B_f}{C_f} - \frac{4A_f}{C_f}}
\]
in eq.(\ref{solpertf}) always lies in the interval $(-1,0)$. Accordingly, the perturbation $\delta f_{\text{hom}}(x)$ diverges as $x \to 0$, even though the singularities of the Tricomi functions have been removed. The only way to cancel this divergence is to impose the condition
\[
j_f! \, c_{1f} + (-1)^{j_f} c_{2f} = 0 \, .
\]
However, it is important to stress that the behavior at $x = 0$ does not affect the physical interpretation of the solutions. As discussed at the beginning of this appendix, a divergence at $x = 0$ for $\delta f$ is in fact expected on general grounds. At $x = 0$, the scaling solution $f = f_* x$ reduces to $f = 0$, which does not admit a physical interpretation, since the effective Newton coupling $g(x) = 1/f(x)$ becomes ill-defined. Accordingly, the parametrization in terms of $f(x)$ breaks down at $x = 0$, and perturbations around this point are not expected to be well-defined. From this perspective, the divergence of $\delta f$ at $x = 0$ is not pathological, but rather signals that this point lies outside the domain where the scaling solution provides a valid description. The physically relevant regime is instead characterized by $x > 0$. In contrast, regularity of the potential $u(x,t)$ at $x = 0$ is still required, since $u$ admits a well-defined physical interpretation in this limit. The different behavior of $u$ and $f$ therefore reflects the fact that the truncation treats the scalar potential and the gravitational coupling in parametrizations with distinct domains of validity.

The solution for $\delta u$ in eq.(\ref{soluvinc}) can be used to clarify the origin of the fixed-point value $x_{0*}$ in Eq.~(\ref{FPbello}). The marginal solution corresponding to $\theta_u = 0$ acts as an effective deformation of the scaling solution. Taking the derivative of $\delta u_{j_u=2}(x,t)$ and setting it to zero reproduces exactly the value of $x_{0*}$ given in Eq.~(\ref{FPbello}). Expanding $\delta f$ around this value yields the truncation
\[
f = \frac{1}{g_*} + f_{1*}(x - x_{0*}) + f_2 (x - x_{0*})^2 + O\!\left((x - x_{0*})^3\right).
\]
where $g_*$ and $f_{1*}$ reproduce the fixed-point values in eq.(\ref{FPbello}). Including also the relevant directions with $\theta_u \neq 0$, the minimum is shifted to 
\begin{equation}
\label{pertlinx0}
\delta x_0(t) = x_{0*} + c_0 e^{-2t}.
\end{equation}
where $c_0$ is a complicated expression involving the quantities in eq.(\ref{solqu}). Replacing eq.(\ref{pertlinx0}) in the solution for $\delta f$ in eq.(\ref{solfpert}), the general perturbations around $x = \delta x_0(t)$ can then be written as
\[
\delta u(x,t) = \delta u_0(t) + \delta \lambda(t)\,(x - \delta x_0(t))^2 + O\!\left((x - \delta x_0(t))^3\right),
\]
\[
\delta f(x,t) = \frac{1}{\delta g(t)} + \delta f_1(t)\,(x - \delta x_0(t)) + O\!\left((x - \delta x_0(t))^2\right),
\]
where $\delta g(t)$ and $\delta f_1(t)$ are given by
\begin{equation}\begin{split}
&\delta g(t)=a_0+a_1 e^{-2t}+a_2 e^{-\theta_5t}+a_3 e^{-\theta_6t}\\
&\delta f_1(t)=b_1 e^{-\theta_5t}+b_2 e^{-\theta_6t}
\end{split}\end{equation}
where the coefficients are complicated functions involving the quantities in eq.(\ref{solqu}) and eq.(\ref{solqf}). Similar expressions can be found by linearizing the flow equations in Appendix \ref{appbeta}, which have been analyzed in section \ref{sectioncriticalexp}. This motivates the general ansatz used in the main text.

As a final remark, we note that the linearized flow equations in Appendix~\ref{appbeta}, expanded around the fixed point in Eq.~(\ref{FPbello}) and used in Sec.~\ref{sectionfixedpoint}, are coupled, whereas the perturbative flows derived in this appendix are decoupled. This difference originates from the fact that the expansion around $x = x_0(t)$ in the full flow equations Eq.~(\ref{adeqfull}) retains partial information about the full functional form of $u(x,t)$, while the present analysis is restricted to perturbations around the scaling solution. Importantly, the critical manifold is determined solely by the perturbations of the scaling solution and is therefore insensitive to the full functional form of $u(x,t)$. This explains why both approaches yield the same set of critical exponents. However, in the full flow, the perturbations of $u_0$, $x_0$, $m_\sigma$, and related quantities also contain contributions involving superpositions of the non-trivial critical exponents associated with $f$, which are absent in the simplified analysis presented in this appendix.
\end{appendices}

\bibliography{reflanpole.bib}
\end{document}